\documentclass{aa}
\usepackage{rotating}
\usepackage{graphicx}
\usepackage{txfonts}
\begin{document}

\title{A test for the theory of colliding winds: the periastron passage of 9~Sagittarii \thanks{Based on observations with XMM-Newton, an ESA Science Mission with instruments and contributions directly funded by ESA Member states and the USA (NASA). Also based on observations collected at the European Southern Observatory (La Silla, Chile) and with the Mercator Telescope operated on the island of La Palma by the Flemish Community, at the Spanish Observatorio del Roque de los Muchachos of the Instituto de Astrof\'{\i}sica de Canarias. Based on observations obtained with the HERMES spectrograph, which is supported by the Fund for Scientific Research of Flanders (FWO), Belgium, the Research Council of K.U.Leuven, Belgium, the Fonds National de la Recherche Scientifique (F.R.S.-FNRS), Belgium, the Royal Observatory of Belgium, the Observatoire de Gen\`{e}ve, Switzerland and the Th\"{u}ringer Landessternwarte Tautenburg, Germany.}}
\subtitle{I. X-ray and optical spectroscopy}
\author{G.\ Rauw\inst{1}  \and R.\ Blomme\inst{2} \and Y.\ Naz\'e\inst{1}\fnmsep\thanks{Research Associate FRS-FNRS (Belgium)} \and M.\ Spano\inst{3} \and L.\ Mahy\inst{1}\fnmsep\thanks{Postdoctoral Researcher FRS-FNRS (Belgium)}  \and E.\ Gosset\inst{1}\fnmsep\thanks{Senior Research Associate FRS-FNRS (Belgium)} \and D.\ Volpi\inst{2} \and \newline H.\ van Winckel\inst{4} \and G.\ Raskin\inst{4} \and C.\ Waelkens\inst{4}}
\offprints{G.\ Rauw}
\mail{rauw@astro.ulg.ac.be}
\institute{Groupe d'Astrophysique des Hautes Energies, Institut d'Astrophysique et de G\'eophysique, Universit\'e de Li\`ege, All\'ee du 6 Ao\^ut, 19c, B\^at B5c, 4000 Li\`ege, Belgium \and Royal Observatory of Belgium, Ringlaan 3, 1180 Brussel, Belgium \and Observatoire de Gen\`eve, Universit\'e de Gen\`eve, 51 Chemin des Maillettes, 1290, Sauverny, Switzerland \and Instituut voor Sterrenkunde, Katholieke Universiteit Leuven, Celestijnenlaan 200\,D, 3001 Leuven, Belgium}
\date{}
\abstract{The long-period, highly eccentric O-star binary 9~Sgr, known for its non-thermal radio emission and its relatively bright X-ray emission, went through its periastron in 2013.}{Such an event can be used to observationally test the predictions of the theory of colliding stellar winds over a broad range of wavelengths.}{We have conducted a multi-wavelength monitoring campaign of 9~Sgr around the 2013 periastron. In this paper, we focus on X-ray observations and optical spectroscopy.}{The optical spectra allow us to revisit the orbital solution of 9~Sgr and to refine its orbital period to 9.1\,years. The X-ray flux is maximum at periastron over all energy bands, but with clear differences as a function of energy. The largest variations are observed at energies above 2\,keV, whilst the spectrum in the soft band (0.5 -- 1.0\,keV) remains mostly unchanged indicating that it arises far from the collision region, in the inner winds of the individual components. The level of the hard emission at periastron clearly deviates from the $1/r$ relation expected for an adiabatic wind interaction zone, whilst this relation seems to hold at the other phases covered by our observations. The spectra taken at phase $0.946$ reveal a clear Fe\,{\sc xxv} line at 6.7\,keV, but no such line is detected at periastron ($\phi = 0.000$) although a simple model predicts a strong line that should be easily visible in the data.} {The peculiarities of the X-ray spectrum of 9~Sgr could reflect the impact of radiative inhibition as well as a phase-dependent efficiency of particle acceleration on the shock properties.}
\authorrunning{Rauw et al.}
\titlerunning{The periastron passage of 9~Sgr}
\maketitle
\section{Introduction \label{intro}}
Massive OB stars host powerful stellar winds driven by their radiation pressure. When two such stars are bound by gravity, these winds interact, giving rise to a number of observational phenomena that span nearly the entire electromagnetic spectrum (for a review see e.g.\ Rauw \cite{Leuven} and Rauw et al.\ \cite{RN}). Two prominent observational diagnostics of wind interactions are strong X-ray emission and synchrotron radio emission. 

Concerning the X-ray emission, observations with the {\it EINSTEIN} and {\it ROSAT} observatories revealed that some of the brightest X-ray sources among massive stars are binary systems (e.g.\ Pollock \cite{Pollock1}, Chlebowski \& Garmany \cite{CG}, Corcoran \cite{Corcoran}). These observations were interpreted in terms of the collision of the supersonic stellar winds producing copious amounts of hot, X-ray emitting plasma (e.g.\ Prilutskii \& Usov \cite{PU}, Cherepashchuk \cite{Chere}, Stevens et al.\ \cite{SBP}). This led to the general paradigm that colliding wind binaries should be bright X-ray emitters. However, over recent years, the detailed studies of large samples of massive stars with well-established multiplicity properties using {\it XMM-Newton} and {\it Chandra} (e.g.\ Sana et al.\ \cite{Sana}, Naz\'e \cite{Naze}, Naz\'e et al.\ \cite{CarinaChandra}, Rauw et al.\ \cite{CygOB2Chandra}) revealed that not all massive binaries are indeed X-ray luminous. For instance, it turned out that O-star binaries with orbital periods of a few days are not particularly bright X-ray sources. This could be due to radiative braking (Gayley et al.\ \cite{Gayley}) and/or radiative inhibition (Stevens \& Pollock \cite{SP}) that lower the pre-shock wind velocities (e.g.\ Pittard \cite{Pittard} and references therein). 

In the radio domain, Abbott et al.\ (\cite{ABC,ABCT}) reported on a sample of O and Wolf-Rayet stars displaying a non-thermal, synchrotron, radio emission in addition to the thermal free-free emission produced in their stellar winds. While the origin of this phenomenon was not clear at the beginning, dedicated multi-wavelength investigations of these objects have shown that they are mostly binary or higher multiplicity systems (e.g.\ Dougherty \& Williams \cite{DW}, van Loo et al.\ \cite{Sven}, Naz\'e et al.\ \cite{YN08,YN10}, Sana et al.\ \cite{HD93250}). The most likely scenario is thus that diffusive shock acceleration of electrons occurs at the shock front produced by the collision of the stellar winds of  massive binary systems (Pittard \& Dougherty \cite{PD} and references therein). 

The bright ($V = 5.97$, $B-V = 0$) star 9~Sgr (= HD~164\,794) is one of the few O-star binaries where both phenomena can be studied. In fact, 9~Sgr was among the first O-type stars discovered to display a synchrotron radio emission (Abbott et al.\ \cite{ABC}) and the star has been known to be an X-ray source since the epoch of the {\it EINSTEIN} satellite. For many years, this star was considered as presumably single. The situation only changed when its X-ray properties as well as episodic variations of some optical line profiles over long time-scales were interpreted as signatures of a long-period (8 - 9 years) eccentric double-lined spectroscopic binary (SB2) system (Rauw et al.\ \cite{PaperI}, hereafter Paper I; Rauw et al.\ \cite{jenam}). This scenario was confirmed when we obtained a first orbital solution of 9~Sgr (Rauw et al.\ \cite{Paper2}, hereafter Paper II) and when phase-locked variations of the radio flux at 2\,cm were reported by Blomme \& Volpi (\cite{BV}, hereafter Paper III). In Paper II, we were able to disentangle the spectra of the primary and secondary components of 9~Sgr, and assigned spectral types O3.5\,V((f$^*$)) and O5-5.5\,V((f)) for the primary and secondary, respectively. 

In 2013, 9~Sgr went through periastron passage. As this is a rather rare event, we have set up a multi-wavelength campaign, including monitoring in the radio, optical and X-rays. In the present paper, we provide the analysis of the new optical and X-ray data. We notably revise the orbital elements of 9~Sgr and discuss the phase-dependence of the X-ray fluxes and spectra. In a forthcoming paper (Blomme et al.\ in prep.), we will present the results of the radio monitoring. 

\begin{table*}[t!]
\caption{Journal of our new optical spectra of 9\,Sgr \label{RVtab}}
\begin{center}
\begin{tabular}{l c c c c c c c}
\hline
HJD-2\,450\,000 & Exp.\ time & Instrument & RV$_1$ & $\sigma_1$ & RV$_2$ & $\sigma_2$ \\
    & (s)        &            & (km\,s$^{-1}$) & (km\,s$^{-1}$) & (km\,s$^{-1}$) & (km\,s$^{-1}$)\\
\hline
4995.5949 &  360 & H & 10.6 & 19.3 & 14.6 & 3.7 \\      
5017.4735 & 1200 & H & 10.5 &  9.3 & 13.2 & 1.9 \\
5037.4475 &  720 & H & 15.4 & 10.1 & 12.7 & 4.0 \\
5080.3962 & 2500 & H & 13.7 &  2.4 & 13.0 & 4.1 \\
5101.3382 &  360 & H & 16.6 &  6.1 & 12.5 & 2.1 \\
5291.7128 &  900 & H & 15.3 &  6.8 & 14.1 & 2.3 \\
5334.6592 &  800 & H & 14.0 &  4.2 & 10.9 & 1.4 \\
5342.6792 & 1200 & H & 13.8 &  4.3 & 10.5 & 2.0 \\
5377.4840 &  900 & H & 15.0 &  8.6 &  9.6 & 2.8 \\
5430.3849 &  960 & H & 12.5 &  6.4 &  9.0 & 2.8 \\
5660.7078 &  600 & H & 18.3 &  5.7 &  6.6 & 2.8 \\
5684.6663 &  600 & H & 16.8 &  9.4 &  6.6 & 1.8 \\
5703.6817 & 2160 & H & 13.9 &  1.7 & 6.9 & 2.6 \\
5704.6231 & 2160 & H & 16.1 &  3.4 & 6.4 & 2.2 \\
5705.5434 &  720 & H & 19.6 & 12.9 & 4.9 & 5.7 \\
5748.5052 &  360 & H & 24.6 &  2.6 & 4.9 & 2.2 \\
5750.4500 &  360 & H & 21.1 & 10.0 & 3.9 & 3.1 \\
5769.5233 & 1200 & H & 16.2 &  7.2 & 5.6 & 1.5 \\
5790.4075 &  460 & H & 21.0 &  4.8 & 1.6 & 5.6 \\
5795.4051 &  500 & H & 18.9 &  2.1 & 4.9 & 3.0 \\
6038.7276 & 1080 & H & 23.8 &  4.0 &$-1.6$ & 2.5 \\
6052.9305 & 300 & F & 22.0 & 18.5 & $-1.9$ & 4.6 \\ 
6053.6602 &  360 & H & 26.1 &  4.1 &$-1.0$ & 5.1 \\
6069.6341 &  360 & H & 26.8 &  8.9 &$-5.7$ & 6.2 \\
6111.5867 &  360 & H & 26.4 &  3.6 &$-3.3$ & 4.4 \\
6125.4789 &  360 & H & 25.8 &  6.2 &$-5.7$ & 4.0 \\
6126.4975 &  600 & H & 25.0 &  7.7 &$-3.7$ & 5.0 \\
6146.4494 &  360 & H & 20.9 & 7.5 &$-12.6$ &11.1 \\
6171.5517 & 600 & C & 27.5 & 3.7 & $-13.9$ & 4.3 \\
6218.4937 & 650 & C & 32.5 & 8.3 & $-16.2$ & 2.2 \\
6354.8983 & 623 & C & 31.3 & 8.3 & $-31.6$ & 6.2 \\
6420.6492 &  480 & H & 46.1 &16.5 &$-45.5$ &10.8 \\
6431.8842 & 623 & C & 40.4 & 13.0 & $-48.4$ & 4.1\\
6458.6064 & 600 & C & 56.2 & 11.2 & $-53.9$ & 4.6\\
6461.5358 &  360 & H & 52.9 &19.2 &$-53.4$ &10.2 \\
6463.5440 &  360 & H & 50.9 &17.7 &$-52.9$ &10.0 \\
6508.6147 & 623 & C & 54.5 & 11.9 & $-57.4$ & 6.1\\
6513.3831 &  690 & H & 57.7 & 8.7 &$-53.8$ & 9.5 \\
6544.5270 & 600 & C & 46.7 & 10.8 & $-54.1$ & 8.0\\
6808.6165 &  360 & H &  9.3 & 5.6 &  7.7 & 8.0 \\
6829.5383 &  360 & H &  5.7 & 1.7 & 12.4 & 2.4 \\
6846.4309 &  360 & H & 17.6 &19.2 & 12.6 & 1.8 \\
6878.4413 &  360 & H & 11.7 & 5.0 & 12.7 & 1.1 \\
6879.4112 &  360 & H &  9.3 & 7.3 & 13.0 & 2.0 \\
6900.3964 &  620 & H & $-1.4$ &11.0 & 14.2 & 1.4 \\
7156.6508 & 1800 & H &  6.0 & 9.8 & 17.3 & 3.4 \\ 
\hline
\end{tabular}
\tablefoot{The codes for the instrument correspond to HERMES (H), FEROS (F) and Coralie (C). The columns labelled $\sigma_1$ and $\sigma_2$ provide the 1-$\sigma$ dispersion about the mean RV of the various lines (corrected for systematic shifts) used to evaluate the RV of the primary and secondary, respectively.\label{journalspecopt}}
\end{center}
\end{table*}

\section{Observations \label{observations}}
As part of our multiwavelength observing campaign to monitor 9~Sgr around its 2013 periastron passage, we have obtained X-ray observations with {\it XMM-Newton} and optical spectroscopy with various instruments.
\subsection{Optical spectra}
New optical spectroscopy of 9~Sgr was gathered with three different instruments (see Table\,\ref{journalspecopt}). 

A total of 38 spectra were obtained between June 2009 and May 2015 with the High Efficiency and Resolution Mercator Echelle Spectrograph (HERMES, Raskin et al.\ \cite{Raskin}) at the 1.2\,m Mercator Telescope at the Roque de los Muchachos Observatory (La Palma, Canary Islands, Spain). 
The resolving power of our observations was R $\simeq$ 85\,000 and the data cover the wavelength range between 3770 and 9000\,\AA. The data were reduced using the standard HERMES pipeline\footnote{http://hermes-as.oma.be/}. These operations include spectral order extraction, flat-fielding and wavelength calibration based on Th-Ar lamp spectra. Cosmic-ray removal, order merging and the barycentric velocity correction were also applied.

Between August 2012 and September 2013, seven spectra were collected with the Coralie spectrograph mounted on the Swiss 1.2\,m Leonhard Euler Telescope at La Silla observatory (Chile). Coralie is an improved version of the Elodie spectrograph (Baranne et al.\ \cite{Baranne}) covering the spectral range from 3850 to 6890\,\AA\ at a resolving power of R = 55\,000. The data were reduced with the Coralie pipeline. 

One additional spectrum was obtained in May 2012 with the FEROS \'echelle spectrograph (Kaufer et al.\ \cite{Kaufer}) on the MPG/ESO 2.2\,m telescope at La Silla. FEROS has a spectral resolving power of 48\,000 over the spectral domain from 3700 to 9200\,\AA. The data were processed with the FEROS reduction pipeline working under MIDAS. 

The spectra from all three instruments were normalized over the spectral domain from 3900 to 6700\,\AA\ by fitting a spline function through the same set of continuum windows as previously used in Paper II. 

\subsection{X-ray spectra}
9~Sgr was observed four times with {\it XMM-Newton} (Jansen et al.\ \cite{Jansen}). The first observation was taken in March 2001 (see Paper I) near orbital phase $\phi = 0.627$ (see Table\,\ref{Xobs}). In March 2013, September 2013 and March 2014, we obtained three additional pointings, specifically sampling phases around periastron passage (see Table\,\ref{Xobs}). In all four observations, the EPIC-pn instrument (Str\"uder et al.\ \cite{pn}) was operated in full frame mode, whilst the EPIC-MOS (Turner et al.\ \cite{MOS}) cameras were operated in large window mode, except for the first observation, where we chose the full frame mode. For each observation and each EPIC instrument, the thick filter was used to reject optical and UV photons. The two reflection grating spectrometers (RGS, den Herder et al.\ \cite{RGS}) were operated in normal spectroscopy mode. 

All four datasets were processed in a homogeneous way with the Scientific Analysis System (SAS) software version 13.5. We checked the high-energy ($> 10$\,keV) EPIC count rates integrated over the entire field of view for soft proton background flares, but none was found. The EPIC spectra of 9~Sgr were extracted over a circular region of radius 15\,arcsec centered on the 2\,MASS coordinates of the source. The spectra were binned to achieve a minimum signal-to-noise ratio of 3 and an oversampling of the spectral resolution of maximum 5. A crucial aspect with these data is the background subtraction. 9~Sgr is located inside the NGC~6530 cluster, and the field of view is populated by a large number of secondary X-ray sources mostly related to low-mass pre-main sequence stars (see Rauw et al.\ \cite{NGC6530}, Damiani et al.\ \cite{Damiani}). However, the most critical contribution to the background comes from the low-mass X-ray binary GX~9+01 (= Sgr~X-3) which is located outside the field of view at about 1$^{\circ}$ from 9~Sgr. This bright off-axis source produces arcs of straylight due to single reflection of the photons on the hyperbolic part of the Wolter I mirrors (see also Paper I). We thus inspected the images of each observation and carefully selected three circular background regions which are close to 9~Sgr and free of secondary sources and straylight features. The RGS1 and RGS2 first and second order spectra were extracted for each observation in a standard way. 

Lightcurves of 9~Sgr and of the background were extracted for each EPIC instrument over the full (0.4 -- 10.0\,keV), the soft (0.4 -- 1.0\,keV) and the hard (1.0 -- 10.0\,keV) energy bands. These lightcurves were converted into equivalent on-axis, full-PSF count rates using the {\it epiclccorr} command under SAS. We adopted time bins of 200 and 1000\,s. The background-corrected count rates of 9~Sgr range between 0.2 and 0.25\,ct\,s$^{-1}$ for the MOS cameras and between 0.8 and 0.95\,ct\,s$^{-1}$ for the pn instrument. These values are well below the pile-up limits of these instruments.
\begin{table*}[thb]
\caption{Journal of the X-ray observations of 9~Sgr\label{Xobs}}
\begin{center}
\begin{tabular}{c c c c c c c}
\hline
ObsID & JD-2\,450\,000 & $\phi$ & Duration & \multicolumn{3}{c}{EPIC count rates (count\,s$^{-1}$)\,[0.5\,keV - 10\,keV]} \\
\cline{5-7}
& & & (ks) & MOS1 & MOS2 & pn\\
\hline
0008820101 & 1\,977.109 & 0.627 & 19.5 & $0.180 \pm 0.003$ & $0.195 \pm 0.003$ & $0.685 \pm 0.006$ \\
0720540401 & 6\,360.297 & 0.946 & 21.9 & $0.176 \pm 0.003$ & $0.198 \pm 0.003$ & $0.767 \pm 0.007$ \\
0720540501 & 6\,539.484 & 0.000 & 23.7 & $0.208 \pm 0.003$ & $0.224 \pm 0.003$ & $0.839 \pm 0.006$ \\
0720540601 & 6\,721.922 & 0.055 & 24.5 & $0.205 \pm 0.003$ & $0.196 \pm 0.003$ & $0.788 \pm 0.006$ \\ 
\hline
\end{tabular}
\end{center}
\end{table*}
\section{A new orbital solution\label{orbital}}
We measured the radial velocities (RVs) in the same way as in Paper II. For spectral lines belonging to only one component of the binary system, we fitted a single Gaussian profile. Lines displaying a clear SB2 signature were fitted with two Gaussians. 

Following Paper II, we then computed the primary RV as the mean of the RVs of the Si\,{\sc iv} $\lambda$\,4116 and N\,{\sc v}\,$\lambda\lambda$\,4604, 4620 lines, whilst the secondary RV was obtained from the mean of the RVs of the secondary components in He\,{\sc i} $\lambda\lambda$\,4471, 5876, O\,{\sc iii} $\lambda$\,5592, C\,{\sc iii}\,$\lambda$\,5696 and C\,{\sc iv}\,$\lambda\lambda$\,5801, 5812. The uncertainties on these RV measurements were evaluated as the 1-$\sigma$ dispersion about these means for each star, after correcting for systematic shifts between the radial velocities of the different lines. The resulting RVs are listed in Table\,\ref{RVtab}. 

We then combined our new RV data with those of Paper II. Doing so, we immediately noted that the 8.6\,year period derived in Paper II was too short. Indeed, according to the ephemerides of Paper II, the partial line deblending, characteristic of periastron passage, should have occurred in March 2013. Our data instead favour a periastron passage in early September 2013.   

\begin{figure}
\resizebox{9cm}{!}{\includegraphics{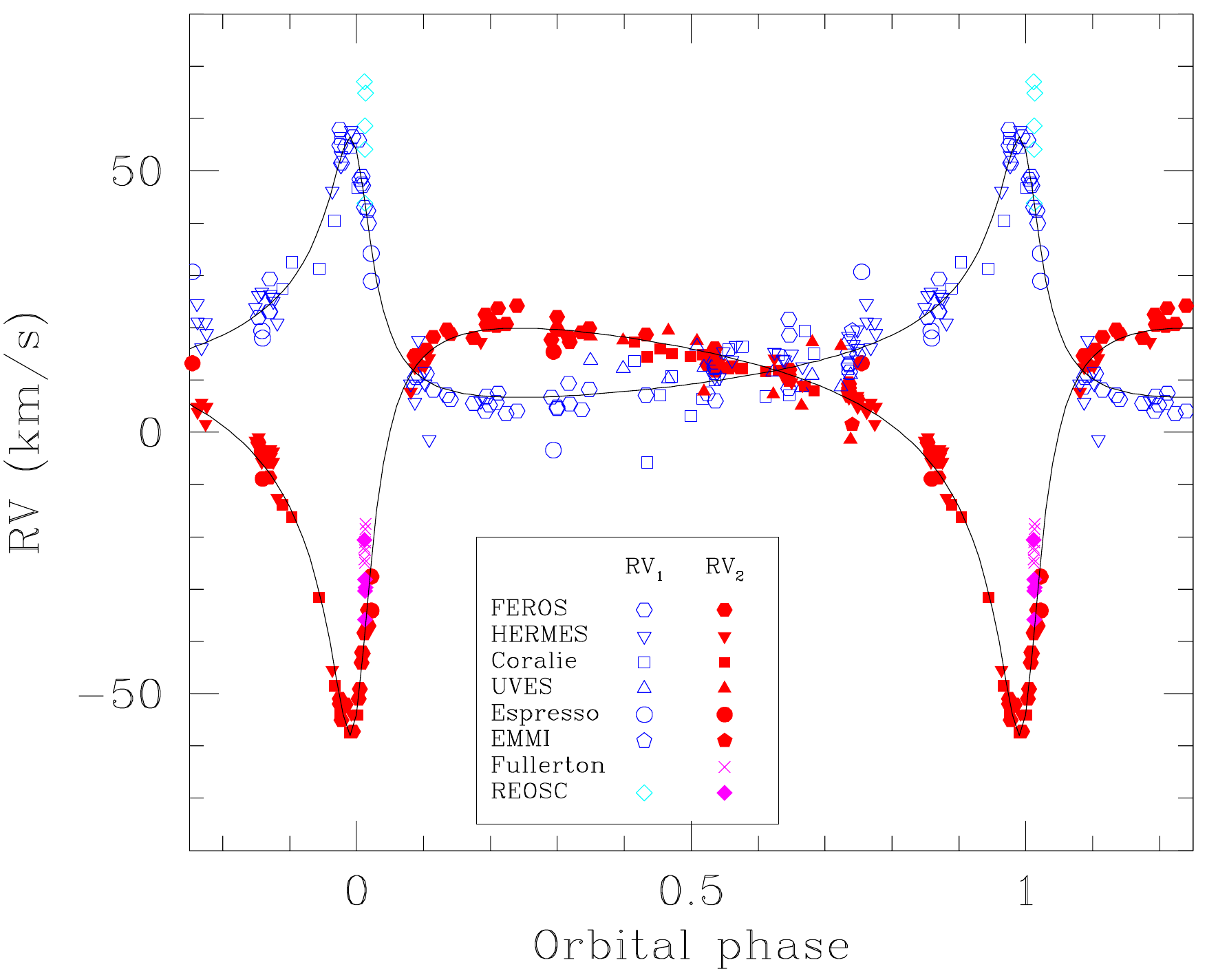}}
\resizebox{9cm}{!}{\includegraphics{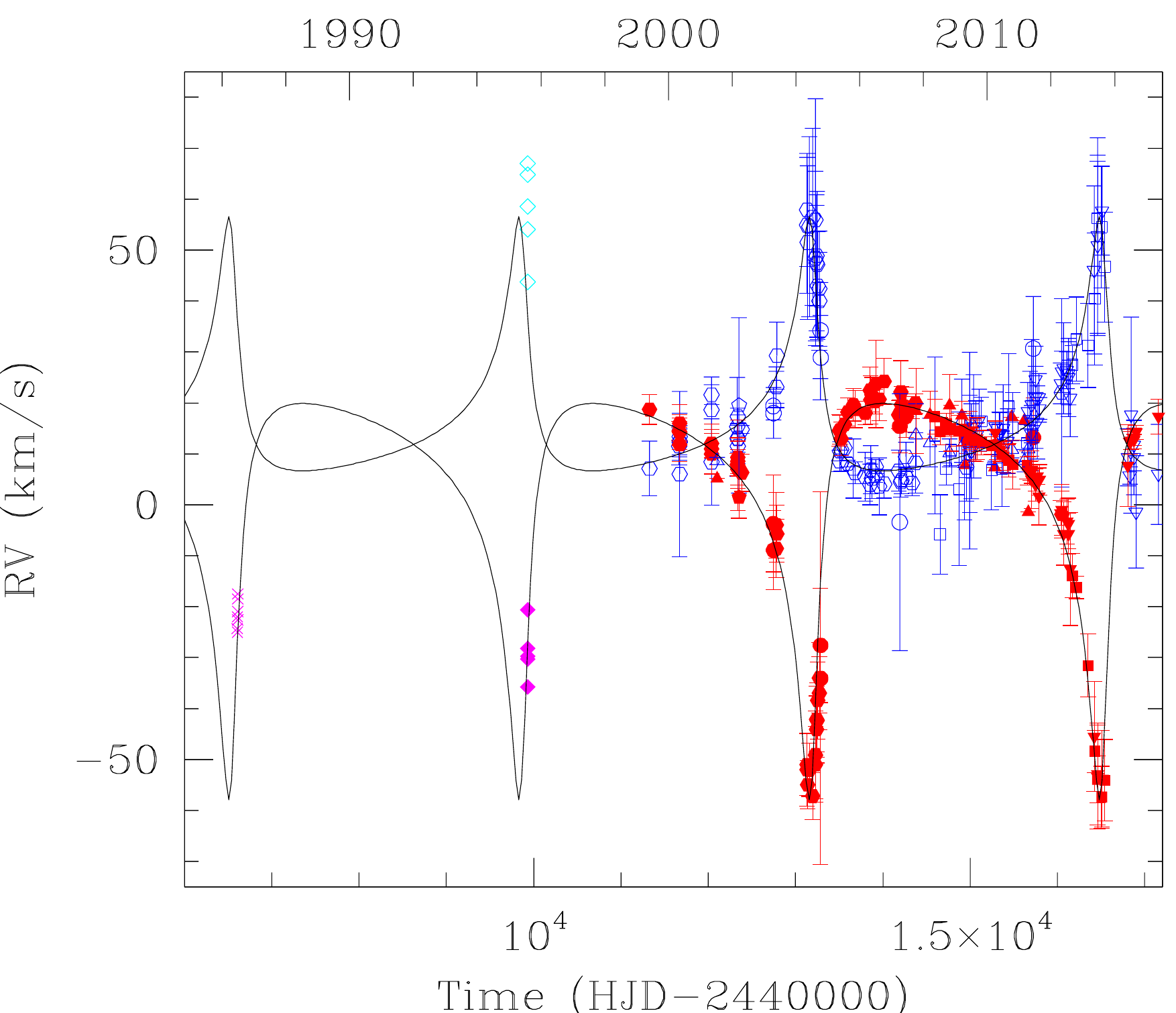}}
\caption{Radial velocity curve of 9~Sgr. Open and filled symbols stand for the RVs of the primary and secondary, respectively. The various symbols refer to data obtained with different instruments. The RVs from Fullerton (\cite{Alex}) refer only to the secondary star and were thus not included in the computation of the SB2 orbital solution. The top panel shows the RV curve folded in orbital phase, whilst the lower panel illustrates the distribution as a function of time (given in years on the top axis and as HJD $-$ 2\,440\,000 on the bottom axis).\label{solsb2}}
\end{figure}

We thus redetermined the orbital period using the generalized Fourier periodogram technique proposed by Heck et al.\ (\cite{HMM}) and amended by Gosset et al.\ (\cite{Gosset}), as well as the trial period method of Lafler \& Kinman (\cite{LK}). We applied these methods to various combinations of the data (secondary RVs, primary RVs minus secondary RVs). In this way, we obtained a best estimate of the orbital frequency of $(3.016 \pm 0.034) \times 10^{-4}$\,d$^{-1}$, corresponding to a period of 3315\,days. We then used the new RVs and those from Paper II to compute an SB2 orbital solution with the LOSP (Li\`ege Orbital Solution Package) code 
 (Sana et al.\ \cite{SGR}) which is an improved version of the code originally proposed by Wolfe et al.\ (\cite{WHS}). We used our above estimate of the orbital period as a starting point and allowed LOSP to iterate on this parameter. The best estimate of the orbital period obtained in this way was $3324.1 \pm 4.3$\,days (i.e.\ 9.1\,years). The parameters inferred from this new orbital solution are listed in Table\,\ref{solorb} and the revised orbital solution is shown in Fig.\,\ref{solsb2}. 

\begin{table}[h!]
\caption{Orbital solution of 9\,Sgr \label{solorb}}
\begin{center}
\begin{tabular}{l c}
\hline
& All SB2 data  \\
\hline
Period (days)       & $3324.1 \pm 4.3$ \\
$T_0$ (HJD$-$2\,450\,000) & $6540.0 \pm 4.4$    \\
$e$                 & $0.705 \pm 0.007$ \\
$\omega$ ($^{\circ}$) & $26.0 \pm 0.9$   \\
$K_1$ (km\,s$^{-1}$) & $24.9 \pm 0.4$   \\
$K_2$ (km\,s$^{-1}$) & $39.0 \pm 0.7$   \\
$\gamma_1$ (km\,s$^{-1}$) & $15.8 \pm 0.3$ \\
$\gamma_2$ (km\,s$^{-1}$) &  $5.6 \pm 0.3$ \\
$a\,\sin{i}$ (R$_{\odot}$) & $2977 \pm 43$  \\
$q = m_2/m_1$           & $0.64 \pm 0.02$  \\
$m_1\,\sin^3{i}$ (M$_{\odot}$) & $19.6 \pm 1.0$ \\
$m_2\,\sin^3{i}$ (M$_{\odot}$) & $12.5 \pm 0.6$ \\
rms(O$-$C)$_1$ (km\,s$^{-1}$) & 4.0 \\
rms(O$-$C)$_2$ (km\,s$^{-1}$) & 2.8 \\
\hline
\end{tabular}
\tablefoot{$T_0$ is the time of periastron passage, $\omega$ is the primary star's longitude of periastron measured from the ascending node of the orbit. The quoted uncertainties correspond to 1-$\sigma$.}
\end{center}
\end{table}

Apart from the orbital period (and hence $a\,\sin{i}$) and the time of periastron passage, the new orbital solution is nearly identical to that of Paper II. We note that the error bars are also smaller than those of the old solution. The reason for the longer period becomes immediately clear when comparing the lower panel of Fig.\,\ref{solsb2} with Fig.\,2 of Paper II. Indeed, in the old orbital solution the REOSC and Fullerton data were erroneously assigned to the descending branch of the secondary's RV curve whilst they now appear on the ascending branch. With two consecutive periastron passages observed with modern high-resolution \'echelle spectrographs, the orbital period is now much better constrained than in Paper II. We note that the revision of the orbital period has no impact on the spectral disentangling performed in Paper II and hence all the results regarding the spectral classifications, rotational velocities, and brightness ratio thus remain unchanged.\\

The large value of $a\,sin{i}$ makes 9~Sgr an interesting target for high angular resolution observations. Indeed, there are two reports in the literature of the system being resolved. Sana et al.\ (\cite{VLTI}) resolved 9~Sgr using the PIONIER instrument on the {\it VLTI} interferometer. Their observation was taken on 2013.2495 ($\phi = 0.953$), i.e.\ only six months before periastron passage. They found $\rho = (4.95 \pm 1.05)$\,mas, $\theta = (242.22 \pm 19.95)^{\circ}$ and $\Delta m_H = 0.45 \pm 0.05$ (or $F_1/F_2 = 1.51 \pm 0.07$, in excellent agreement with our determination of the optical brightness ratio $F_1/F_2$ in the range $1.0$ -- $1.5$ in Paper II). Aldoretta et al.\ (\cite{Aldoretta}) report an observation taken on 2008.1920 ($\phi = 0.397$) with the Fine Guidance Sensor (FGS) onboard {\it HST}. Since the 9~Sgr system is only partially resolved with the FGS, there is an ambiguity between the separation $\rho$ and the optical brightness ratio $F_1/F_2$, and a four-fold ambiguity on the position angle $\theta$. Aldoretta et al.\ (\cite{Aldoretta}) report $\rho_{\rm min} = (18.5 \pm 1.9)$\,mas assuming a brightness ratio unity. Assuming $F_1/F_2 = 1.5$ (Paper II, Sana et al.\ \cite{VLTI}), equation (4) of Aldoretta et al.\ (\cite{Aldoretta}) then allows us to compute $\rho = (18.9 \pm 1.9)$\,mas.

\begin{figure}
\resizebox{9cm}{!}{\includegraphics{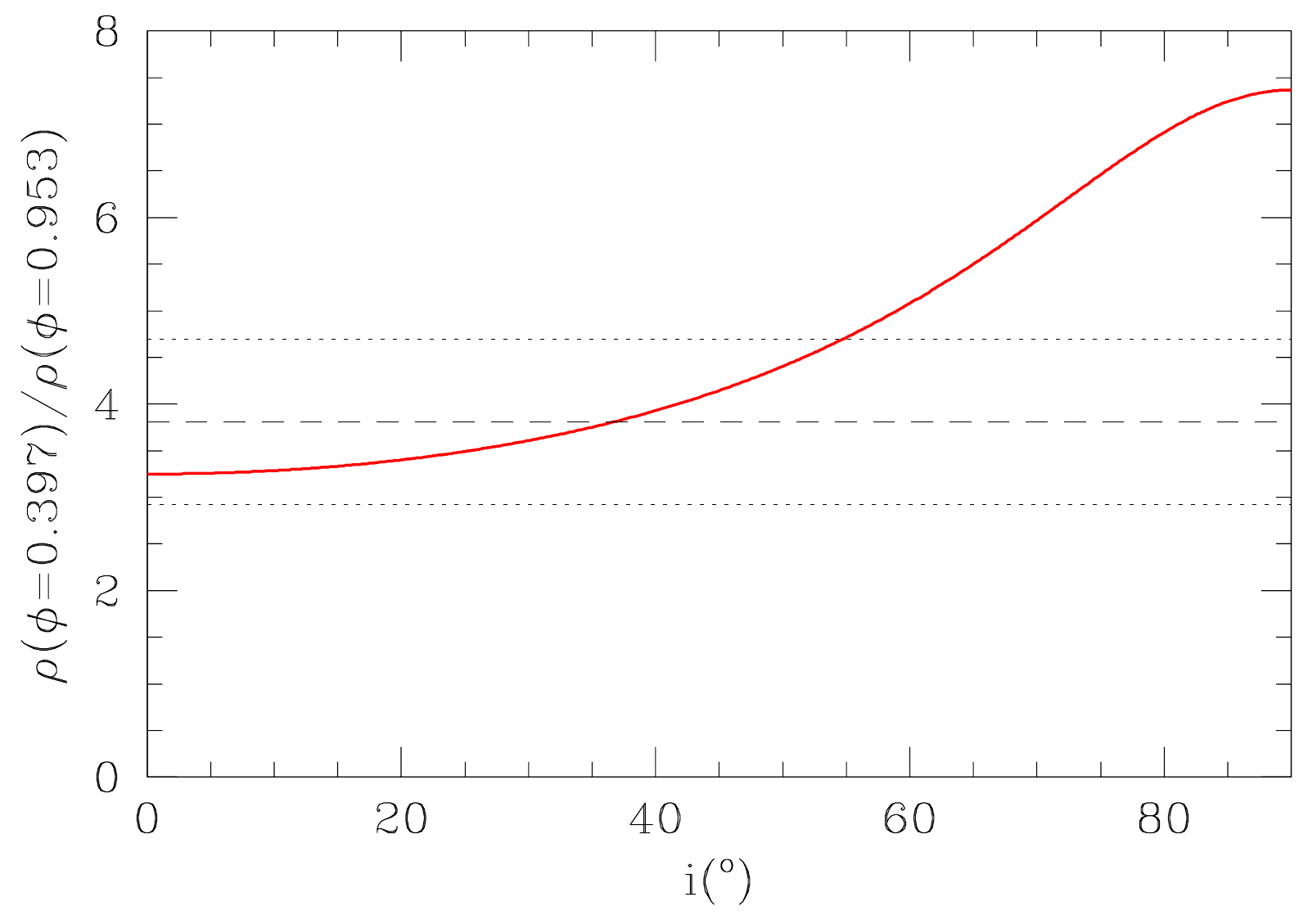}}
\caption{Ratio between the angular separation of the stars in 9~Sgr at orbital phases $0.397$ and $0.953$ as a function of orbital inclination (red solid curve). The dashed line yields the observed ratio and the dotted lines yield the corresponding error interval.\label{inclin}}
\end{figure}

The observed ratio $\rho(\phi=0.397)/\rho(\phi=0.953) = 3.81 \pm 0.89$ can be used to constrain the inclination of the orbit of 9~Sgr. Indeed, this ratio depends upon the longitude of periastron, the eccentricity, the true anomalies at the times of the observations and the orbital inclination. The first two parameters are known from our orbital solution (Table\,\ref{solorb}), and the third one can be directly computed from the orbital phase and our orbital solution. Therefore, the only remaining free parameter is the orbital inclination. This comparison is shown in Fig.\,\ref{inclin}. As can be seen from this figure, the current best value of $i$ would be about $37^{\circ}$. If we account for the error bars on the ratio of the angular separations, we find an upper limit on $i$ of $55^{\circ}$. We note that these numbers are in very reasonable agreement with the estimate of the inclination of $(45 \pm 1)^{\circ}$ obtained by comparing the minimum masses from Table\,\ref{solorb} with the spectroscopic masses quoted by Martins et al.\ (\cite{Martins}). An inclination of $37^{\circ}$ would imply a distance of 1.985\,kpc, slightly larger than the distance of the NGC~6530 cluster of $(1.78 \pm 0.08)$\,kpc (Sung et al.\ \cite{Sung}) and our spectroscopic distance of 9~Sgr (1.79\,kpc) derived in Paper II. The same inclination would lead to very large, though not impossible, masses (90 and 58\,M$_{\odot}$ for the primary and secondary, respectively). However, as stated above, accounting for the uncertainties on the ratio of angular separations easily allows us to obtain more reasonable masses. 

A more intensive interferometric monitoring has been performed over recent years with the PIONIER instrument on the {\it VLTI}. Quite surprisingly, a preliminary analysis of these data suggests an orbital inclination close to $90^{\circ}$ (H.\ Sana, private communication). Combined with our orbital solution, such an inclination would imply very low stellar masses and would further suggest that 9~Sgr be a foreground object compared to the NGC~6530 cluster. The origin of this discrepancy is currently unclear.

\section{X-ray spectroscopy}
\begin{figure}
\resizebox{9cm}{!}{\includegraphics{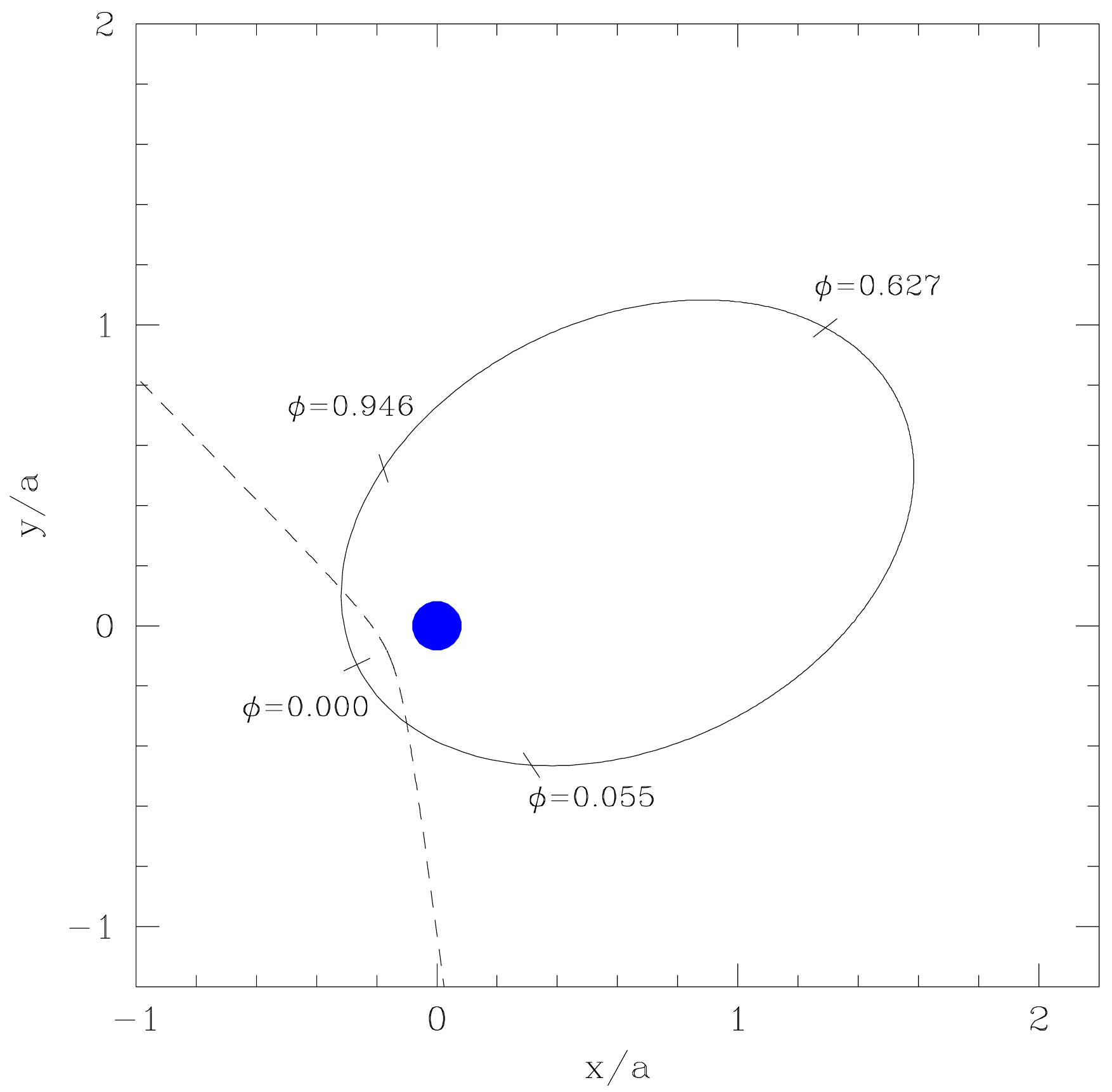}}
\caption{Orbital plane of 9~Sgr. The ellipse shows the orbit of the secondary star about the primary (shown as the big blue dot at the origin of the axes). The tickmarks indicate the positions of the secondary at the times of our {\it XMM-Newton} observations. The dashed line corresponds to the contact surface between the stellar winds at periastron, computed according to the formalism of Cant\'o et al.\ (\cite{Canto}). The projection of the Earth in this plane is located at $(0,-D/a\,\sin{i})$ where $D$ is the distance of 9~Sgr.\label{ellipse}}
\end{figure}

Our {\it XMM-Newton} observations sample one orbital phase close to apastron and three phases around periastron (see Table\,\ref{Xobs} and Fig.\,\ref{ellipse}). If the system hosts a wind interaction that produces X-ray emission, we expect to observe changes in the X-ray spectra between the various phases. These changes can result from a varying line-of-sight column density or from the changing separation between the stars or from a combination of both effects (Stevens et al.\ \cite{SBP}, Pittard \& Parkin \cite{PP}).

\begin{figure}
\resizebox{9cm}{!}{\includegraphics{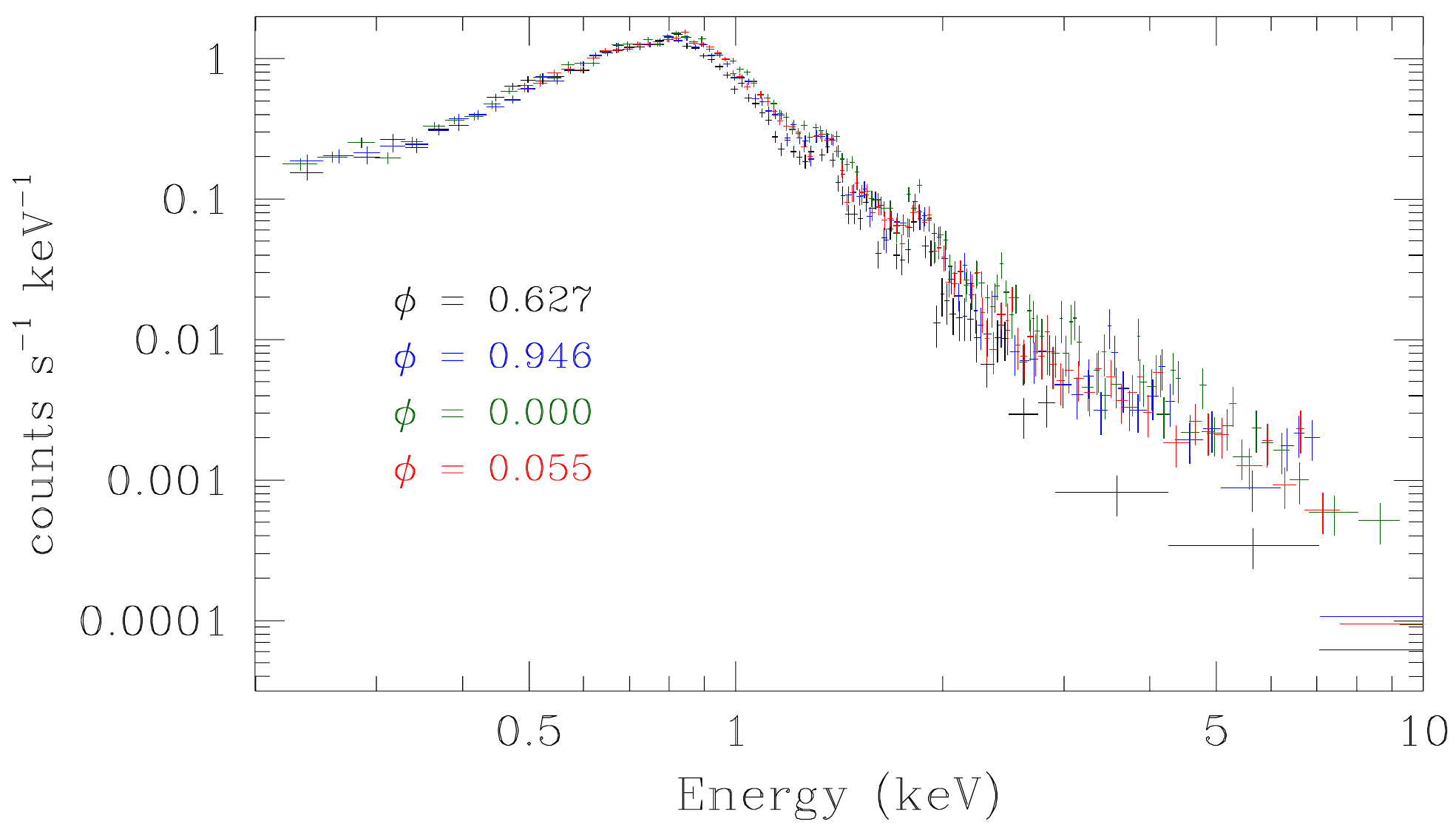}}
\caption{Comparison of the EPIC-pn spectra at four different orbital phases of 9~Sgr. We caution that the variations of the instrumental sensitivity, as the mission gets older, could bias this comparison. The data at $\phi = 0.627$ were taken 12 -- 13 years before those at the other phases, and thus probably benefit from a higher sensitivity of the instrument. The sensitivity likely remained essentially unchanged between the three other pointings.\label{EPICall}}
\end{figure}

Figure\,\ref{EPICall} illustrates the EPIC-pn spectra of 9~Sgr at the four epochs that have been monitored. This figure reveals surprisingly little variability of the X-ray emission, especially at energies below 0.9\,keV. However, the hard tail of the spectrum, which is quite weak and steeply falls-off at $\phi = 0.627$, strongly increases and flattens on the three observations taken around periastron passage. A detailed discussion of the spectral variability needs to account for the changing instrumental response due to the ageing of the detector. This effect is lower for the EPIC-pn instrument compared to the EPIC-MOS. Still, when considering Fig.\,\ref{EPICall}, we need to keep in mind that the data at $\phi = 0.627$ were taken 12 -- 13 years before those around periastron passage, and thus probably benefit from a higher sensitivity of the instrument. The variations of the sensitivity are fully accounted for in the response files that we use in our spectral analysis and discussion of the inter-pointing variations in Sect.\,\ref{discussion}. 

We have also investigated the possibility of intra-pointing variability. Following the approach of Naz\'e et al.\ (\cite{zetaPup2}), we performed $\chi^2$ tests of the background- and barycentric-corrected EPIC lightcurves of each observation, keeping only bins with more than 80\% fractional exposure time. These tests were performed for several hypotheses (constancy, linear trend, quadratic trend). No significant ($SL <$ 1\%) intra-pointing variability was detected with these tests for any of the four observations. 

\subsection{High-resolution X-ray spectroscopy \label{highres}}
\begin{figure}
\resizebox{9cm}{!}{\includegraphics{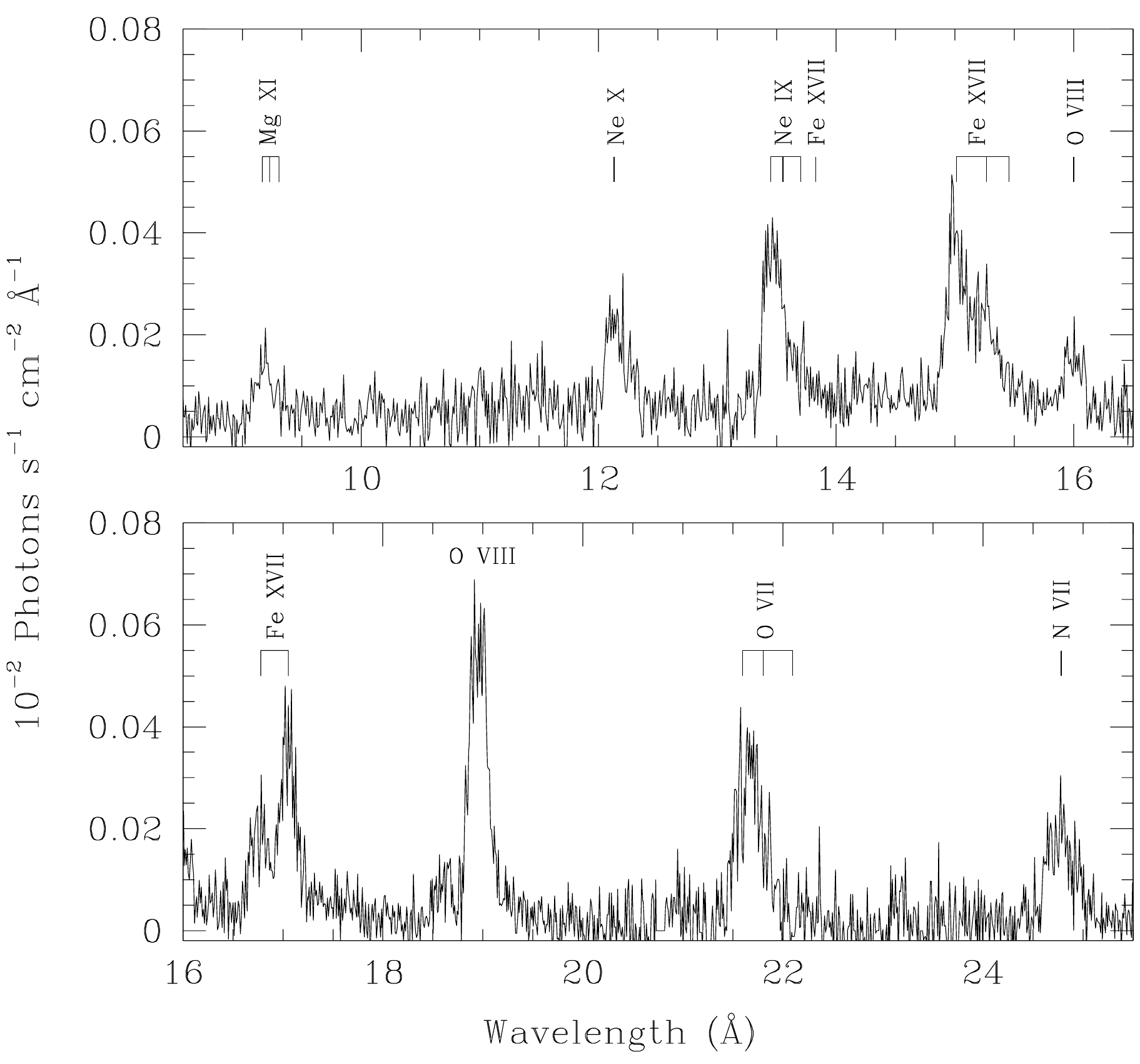}}
\caption{Combined RGS spectrum (including first and second orders of RGS1 and RGS2) of all four X-ray observations. The most important spectral lines are identified by the labels. \label{RGSfluxer}}
\end{figure}

Because 9~Sgr is a rather bright source, and despite the rather short exposure times, high-resolution RGS spectra are available and provide useful information on the X-ray emission in the soft and medium band, at least for the combined RGS data from all four observations. Indeed, the fact that there is little variation of the X-ray spectrum in the soft band (see Fig.\,\ref{EPICall}) allowed us to combine the RGS spectra of the four observations into a single spectrum. The resulting fluxed spectrum is shown in Fig.\,\ref{RGSfluxer}. 

The RGS spectrum of 9~Sgr obtained at $\phi = 0.627$ was already presented and discussed in Paper I. With the fourfold exposure time available now, the new combined RGS spectrum is of significantly better quality, and we can use it to further characterize the soft X-ray emission. Unfortunately, a detailed modelling such as done for $\zeta$~Pup (Herv\'e et al.\ \cite{zetaPup}) or $\lambda$~Cep (Rauw et al.\ \cite{lamCep}) cannot be done in this case. Indeed, even though the emission in the soft band is probably only very little affected by the wind interaction, 9~Sgr hosts two stars, both of which are likely to contribute to the X-ray emission. If we assume that the intrinsic emission of both stars follows the same scaling relation with bolometric luminosity, we can estimate that the primary contributes about twice as much soft X-rays as the secondary star. Therefore, the RGS spectrum cannot be analysed as if it were due to a single star.

Figure\,\ref{RGSfluxer} reveals a number of strong emission lines, clearly indicating that the soft part of the spectrum of 9~Sgr arises in a thermal plasma. The strongest spectral features in the RGS spectrum are several \ion{Fe}{xvii} lines, the Ly$\alpha$ lines of \ion{N}{vii}, \ion{O}{viii} and \ion{Ne}{x}, the Ly$\beta$ line of \ion{O}{viii}, and the He-like triplets of \ion{O}{vii}, \ion{Ne}{ix} and \ion{Mg}{xi}. The temperature of the emitting plasma can be estimated from the ratio of the fluxes of the hydrogen and helium-like ions provided these lines arise in the same region. In our case, this exercise could be performed for oxygen and neon. After correcting the fluxes for the wavelength dependence of the interstellar and wind absorption, using the column densities obtained in our broadband analysis (see the forthcoming Sect.\,\ref{broad}), we inferred values of $1.9$ and $0.49$ for the \ion{Ne}{ix} triplet / \ion{Ne}{x} Ly$\alpha$ and \ion{O}{vii} triplet / \ion{O}{viii} Ly$\alpha$ flux ratios, respectively. These ratios suggest plasma temperatures near $0.3$\,keV. 

The He-like triplets, which consist of a resonance line ($r$), an intercombination doublet ($i$) and a forbidden line ($f$), can be used as a diagnostic of the UV radiation field in the region of the X-ray plasma (for details see Gabriel \& Jordan \cite{GJ}, Blumenthal et al.\ \cite{Blumenthal}, Porquet et al.\ \cite{Porquet}). The RGS spectrum of 9~Sgr indicates that the empirical $f/i$ intensity ratios are strongly reduced compared to the ratios of the transition probabilities. For instance, for the \ion{Ne}{ix} triplet, we obtain $f/i = 0.05^{+.20}_{-.05}$ on the combined RGS spectrum. This value is better constrained than in Paper I because of the higher S/N of the RGS spectrum analysed here. Comparing this ratio with the tables of Porquet et al.\ (\cite{Porquet}) and with the radial dependence of the $f/i$ ratio shown in Fig.\,4 of Paper I indicates that the \ion{Ne}{ix} lines must form in a harsh UV radiation field. Indeed, our best value of the $f/i$ ratio indicates a formation region two stellar radii above photosphere, corresponding to a dilution factor of 0.03. If we account for the uncertainty on the $f/i$ ratio, we obtain an outer limit of eight stellar radii above the photosphere, corresponding to a minimum value of the dilution factor of 0.004. This result clearly indicates that this emission is {\em not} coming from the wind interaction zone. Indeed, even at periastron passage, i.e.\ at the phase where the impact of UV radiation on the plasma in the wind interaction zone is largest, the separation between the two stars amounts to $(881.5/\sin{i})$\,R$_{\odot}$, i.e.\ about $(67/\sin{i})$\,R$_*^p$ or $(83/\sin{i})$\,R$_*^s$. The shock would then be located at about $(42/\sin{i})$\,R$_*^p$ and $(31/\sin{i})$\,R$_*^s$ from the primary and secondary star, respectively (see Sect.\,\ref{discussion}). For our upper limit on the inclination of $55^{\circ}$, these distances correspond to dilution factors of the photospheric UV radiation field of order $2.6 \times 10^{-4}$, i.e. much lower than what we find here from the $f/i$ ratio. We thus conclude that the bulk of the soft X-ray emission of 9~Sgr is indeed intrinsic to the individual stars and does not arise in the wind interaction zone.

Finally, a comparison of the emissivities of the \ion{N}{vii} L$\alpha$ and \ion{O}{viii} L$\alpha$ lines at $kT = 0.3$\,keV against the absorption-corrected flux ratio of these lines suggests an N/O ratio of about $2.7 \pm 0.5$ times solar. 

\subsection{Spectral fitting \label{broad}}
The results from Sect.\,\ref{highres} can be used as a guideline for the broad-band fitting of the EPIC + RGS spectra. Indeed, in the previous section we have found that at least the soft emission is thermal with $kT \sim 0.3$ and arises in a plasma with a slightly enhanced nitrogen abundance.

The X-ray spectra were fitted using the {\tt xspec} v.12.9.0 spectral analysis software package (Arnaud \cite{xspec}). We performed the spectral fitting for the EPIC and RGS spectra together as well as for the sole EPIC spectra. For the EPIC spectra we considered spectral bins between 0.3 and 10.0\,keV, whilst we restricted the RGS data to wavelengths between 9 and 25\,\AA. The spectral fits obtained for the EPIC data alone are of better statistical quality than those obtained when EPIC and RGS spectra are considered simultaneously\footnote{We stress though that the resulting model parameters are very similar.}. As pointed out above, the detailed fitting of the RGS spectra of early-type stars requires sophisticated models to account for the line morphology as well as for the effects of photospheric UV radiation. Because these models are not suited to describe a binary system such as 9~Sgr, we mainly focus on the fits of the EPIC spectra which are much less sensitive to these aspects. Even in this case, the $\chi^2_{\nu}$ values of the best fits are rather large. This is likely due to the fact that for well exposed spectra with small error bars, the systematic uncertainties on the instrumental responses become dominant, but these errors are not properly accounted for in the formal error bars (see e.g.\ Naz\'e et al.\ \cite{YN12b}).

As a first step, we have attempted to fit these spectra with absorbed {\tt vapec} optically-thin thermal plasma models (Smith \& Brickhouse \cite{apec}) with solar abundances (Anders \& Grevesse \cite{AG}), except for nitrogen which was set to 2.7 times solar. 

The absorption arises from the neutral interstellar medium (ISM) and from the ionized stellar winds. For the ISM, we adopted a fixed neutral hydrogen column density of $2.34 \times 10^{21}$\,cm$^{-2}$ corresponding to 1.2 times the H\,{\sc i} column density determined by Diplas \& Savage (\cite{DS}) from the interstellar Ly$\alpha$ absorption towards 9~Sgr. The value of $1.95 \times 10^{21}$\,cm$^{-2}$ quoted by Diplas \& Savage (\cite{DS}), which agrees very well with earlier measurements of Shull \& Van Steenberg (\cite{Shull}), accounts for the column density of atomic hydrogen only. When dealing with the interstellar hydrogen column density, one additionally needs to account for the column density of molecular hydrogen. On average the molecular hydrogen part amounts to $\sim 20$\% of the atomic part (see Table 2 of Bohlin et al.\ \cite{Bohlin}). The full neutral hydrogen column density N(H) = N(H\,{\sc i} + H$_2$) correlates with the $E(B - V)$ colour excess (Bohlin et al.\ \cite{Bohlin}, Gudennavar et al.\ \cite{Gudennavar}). Yet, in the case of 9~Sgr, $B - V = 0.0$ leads to values of N(H) equal to $1.62 \times 10^{21}$ (Bohlin et al.\ \cite{Bohlin}) or $1.71 \times 10^{21}$\,cm$^{-2}$ (Gudennavar et al.\ \cite{Gudennavar}) which are less than the column density of atomic hydrogen. Hence, it seems that the general correlations of Bohlin et al.\ (\cite{Bohlin}) and Gudennavar et al.\ (\cite{Gudennavar}) underestimate the actual neutral column density in the case of 9~Sgr. To account for the contribution of H$_2$ we have thus applied a 20\% correction to the H\,{\sc i} column as determined by Diplas \& Savage (\cite{DS}). For the photoelectric absorption by the interstellar medium, we use the {\tt phabs} model within {\tt xspec} with cross-sections from Ba\l uci\'nska-Church \& McCammon (\cite{BC}) and Yan et al.\ (\cite{Yan}). For the ionized stellar wind, we imported the stellar wind absorption model of Naz\'e et al.\ (\cite{HD108}) into {\tt xspec} as a multiplicative tabular model (hereafter labelled as {\tt mtab\{wind\}}).

A first series of tests was performed with a model of the kind 
\begin{equation}
{\tt phabs}*\left(\sum_{j=1}^{n}{\tt mtab\{wind_j\}*vapec(T_j)}\right)
\label{eq1}
\end{equation}
Single temperature models ($n = 1$) clearly cannot account for the full energy range covered by the spectrum. We thus tested multi-temperature models with $n = 2$ or $3$. Except for the spectrum at $\phi = 0.627$, fitting the EPIC data at high energies required three plasma components. We further noted that the lowest temperature component had very stable properties as one could expect from the constancy of the soft emission (see Fig.\,\ref{EPICall}). To within a few percent, its plasma temperature ($kT_1 = 0.265$\,keV) and normalization parameter ($5.24 \times 10^{-3}$\,cm$^{-5}$), as well as the associated wind column density ($2.94 \times 10^{21}$\,cm$^{-2}$), are all independent of the orbital phase and agree quite well with the results of our analysis of the combined RGS spectrum in Sect.\,\ref{highres}. In the subsequent spectral fits, we thus fixed these model parameters. Our tests further revealed that the wind-column of the highest temperature plasma in Eq.\,\ref{eq1} is essentially unconstrained. As a consequence, we thus adopted the following expression for the thermal plasma model:     
\begin{eqnarray}
& & {\tt phabs}*[{\tt mtab\{wind_1\}*vapec(T_1)} +  \nonumber \\
& & {\tt mtab\{wind_2\}*(vapec(T_2) + vapec(T_3)})] 
\label{eq2}
\end{eqnarray}
In this expression {\tt wind$_1$} had a fixed $N_{\rm wind,1}$ of $2.94 \times 10^{21}$\,cm$^{-2}$, $kT_1$ was set to 0.265\,keV and its normalization parameter to $5.24 \times 10^{-3}$\,cm$^{-5}$. The results of the fits with this model are quoted in the upper part of Table\,\ref{EPICfits}. We note that the wind column density towards the second and third plasma component is most of the time very poorly constrained and the best fit is quite frequently found at the lower limit of $\log{N_{wind,2}}$ which we set at 19.90 in our models.  

\begin{table*}[thb]
\caption{Best fit parameters of the EPIC spectra of 9~Sgr\label{EPICfits}}
\begin{center}
\begin{tabular}{c c c c c c c c c}
\hline
\vspace*{-3mm}\\
\multicolumn{8}{c}{Purely thermal plasma models with ionization equilibrium (Eq.\,\ref{eq2})} \\
\hline
\vspace*{-3mm}\\
$\phi$ & $\log{N_{wind,2}}$ & $kT_2$ & $10^4 \times$ norm$_2$ & $kT_3$ & $10^4 \times$ norm$_3$ & $\chi_{\nu}^2$&  d.o.f. \\ 
       & (cm$^{-2}$)       & (keV)  &(cm$^{-5}$)& (keV) &(cm$^{-5}$) & & \\
\hline
\vspace*{-3mm}\\
0.627 & $21.41^{+.10}_{-.15}$ & $0.73^{+.03}_{-.03}$ & $4.37^{+.57}_{-.57}$ & -- & -- & 1.70 & 229 \\
\vspace*{-3mm}\\
0.946 & $19.90$ & $0.81^{+.03}_{-.03}$ & $3.03^{+.14}_{-.18}$ & $3.57^{+0.79}_{-0.61}$ & $2.05^{+.27}_{-.17}$ & 1.39 & 268 \\
\vspace*{-3mm}\\
0.000 & $19.90$ & $0.82^{+.02}_{-.02}$ & $3.68^{+.14}_{-.17}$ & $2.47^{+0.22}_{-0.20}$ & $3.20^{+.26}_{-.26}$ & 1.72 & 310 \\
\vspace*{-3mm}\\
0.055 & $19.90$ & $0.79^{+.02}_{-.02}$ & $3.30^{+.13}_{-.15}$ & $3.00^{+0.40}_{-0.37}$ & $2.30^{+.23}_{-.16}$ & 1.50 & 293 \\
\vspace*{-3mm}\\
\hline
\vspace*{-3mm}\\
\multicolumn{9}{c}{Thermal models with non-equilibrium ionization (Eq.\,\ref{eq4})} \\
\hline
\vspace*{-3mm}\\
$\phi$ & $\log{N_{wind,2}}$ & $kT_2$ & $10^4 \times$ norm$_2$ & $kT_3$ & $10^4 \times$ norm$_3$ & $10^{-10} \times \tau_u$ & $\chi_{\nu}^2$&  d.o.f. \\ 
       & (cm$^{-2}$)       & (keV)  &(cm$^{-5}$) & (keV) &  (cm$^{-5}$) & (s\,cm$^{-3}$) & & \\
\hline
\vspace*{-3mm}\\
0.627 & $21.15^{+.20}_{-.39}$ & $0.60^{+.09}_{-.13}$ & $1.65^{+.65}_{-.54}$ & $2.9^{+1.3}_{-0.8}$ & $0.88^{+.27}_{-.21}$ & $4.6^{+1.4}_{-1.2}$ & 1.27 & 226 \\
\vspace*{-3mm}\\
0.946 & $19.90$ & $0.78^{+.04}_{-.05}$ & $2.48^{+.17}_{-.52}$ & $7.2^{+.2.5}_{-3.3}$ & $1.41^{+.25}_{-.24}$ & $26^{+116}_{-10}$ & 1.36 & 267 \\
\vspace*{-3mm}\\
0.000 & $20.96^{+.19}_{-.39}$ & $0.75^{+.04}_{-.05}$ & $2.49^{+.52}_{-.47}$ & $6.6^{+.1.5}_{-1.2}$ & $1.89^{+.17}_{-.15}$ & $9.1^{+2.4}_{-1.6}$ & 1.41 & 309 \\
\vspace*{-3mm}\\
0.055 & $19.99^{+.98}_{...}$ & $0.75^{+.03}_{-.04}$ & $2.41^{+.56}_{-.15}$ & $7.5^{+.1.9}_{-2.4}$ & $1.42^{+.15}_{-.14}$ & $15.5^{+20.3}_{-5.9}$ & 1.43 & 292 \\
\vspace*{-3mm}\\
\hline
\vspace*{-3mm}\\
\multicolumn{8}{c}{Models with a power-law component (Eq.\,\ref{eq3})} \\
\hline
\vspace*{-3mm}\\
$\phi$ & $\log{N_{wind,2}}$ & $kT_2$ & $10^4 \times$ norm$_2$ & $\Gamma_3$ & $10^4 \times$ norm$_3$ & $\chi_{\nu}^2$&  d.o.f. \\ 
       & (cm$^{-2}$)       & (keV)  &(cm$^{-5}$)&          &(photons\,keV$^{-1}$\,cm$^{-2}$\,s$^{-1}$) & & \\
\hline
\vspace*{-3mm}\\
0.627 & $19.90$ & $0.74^{+.04}_{-.04}$ & $1.87^{+.19}_{-.19}$ & $3.60^{+.17}_{-.20}$ & $1.35^{+.22}_{-.23}$ & 1.36 & 227 \\
\vspace*{-3mm}\\
0.946 & $19.90$ & $0.81^{+.03}_{-.03}$ & $2.76^{+.15}_{-.22}$ & $2.36^{+.19}_{-.20}$ & $1.21^{+.28}_{-.26}$ & 1.36 & 268 \\
\vspace*{-3mm}\\
0.000 & $19.90$ & $0.91^{+.02}_{-.03}$ & $3.37^{+.24}_{-.24}$ & $2.63^{+.13}_{-.14}$ & $2.04^{+.32}_{-.32}$ & 1.46 & 310 \\
\vspace*{-3mm}\\
0.055 & $19.90$ & $0.81^{+.03}_{-.02}$ & $3.05^{+.14}_{-.20}$ & $2.43^{+.16}_{-.17}$ & $1.33^{+.26}_{-.24}$ & 1.46 & 293 \\
\vspace*{-3mm}\\
\hline
\end{tabular}
\end{center}
\tablefoot{The EPIC spectra were fitted for energies between 0.3 and 10.0\,keV. The interstellar neutral hydrogen column density was fixed to $0.234 \times 10^{22}$\,cm$^{-2}$. Each model further included a soft thermal component with fixed parameters: $\log{N_{\rm wind,1}} = 21.47$, $kT_1 = 0.265$\,keV, norm$_1 = 5.24 \times 10^{-3}$\,cm$^{-5}$. All abundances were set to solar, except for nitrogen which was taken to be 2.7 times solar. The normalization of the thermal models is given as $10^{-14}\,\frac{\int n_e\,n_H\,dV}{4\,\pi\,D^2}$ (cm$^{-5}$) where $n_e$ and $n_H$ are the electron and proton densities of the X-ray emitting plasma in cm$^{-3}$, and $D$ is the distance in cm. The normalization of the power-law component is expressed as the number of photons\,keV$^{-1}$\,cm$^{-2}$\,s$^{-1}$ at 1\,keV. The quoted error bars correspond to the 90\% confidence level.}
\end{table*}

\begin{figure*}
\begin{minipage}{8cm}
\resizebox{8cm}{!}{\includegraphics{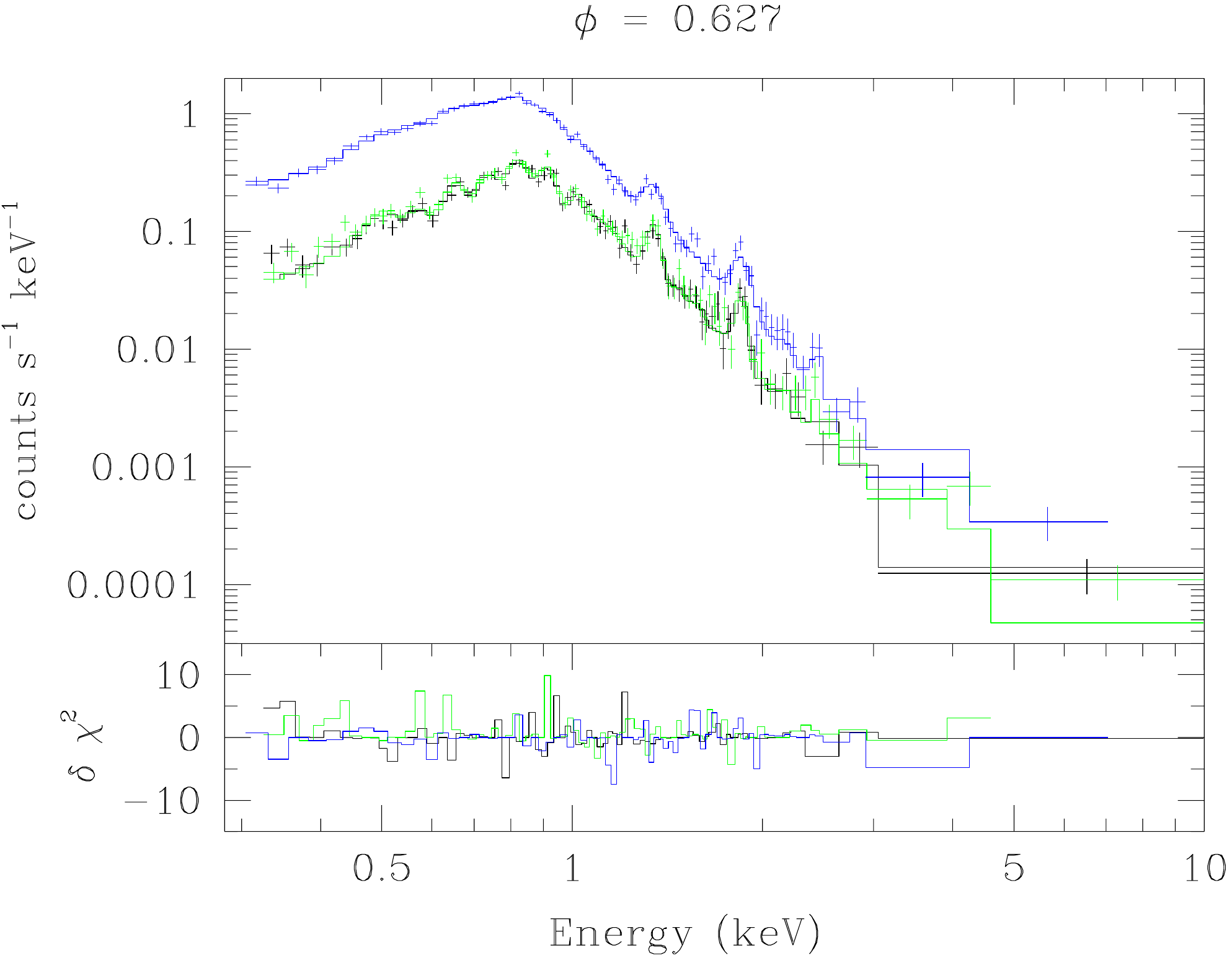}}
\end{minipage}
\begin{minipage}{8cm}
\resizebox{8cm}{!}{\includegraphics{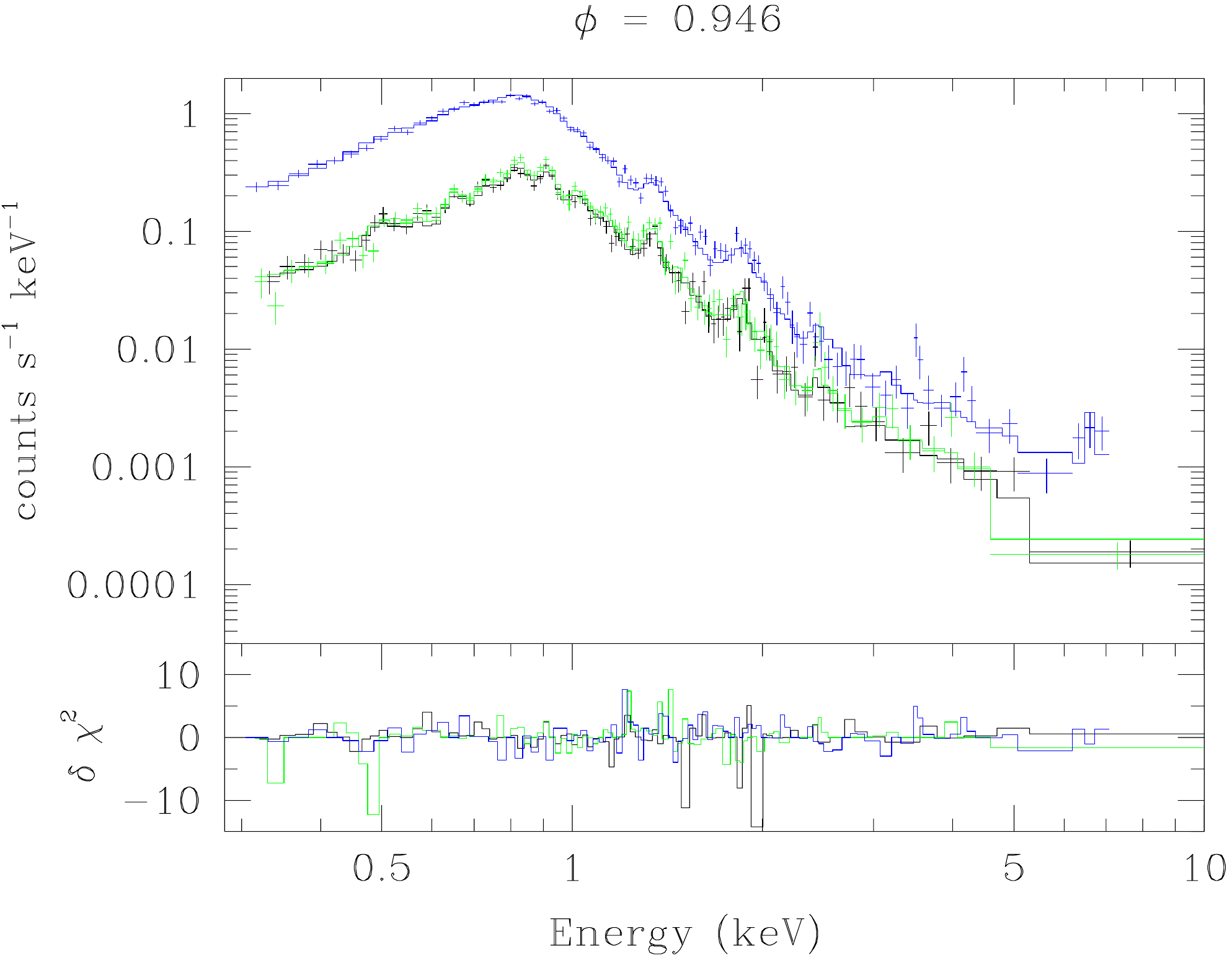}}
\end{minipage}

\vspace*{3mm}
\begin{minipage}{8cm}
\resizebox{8cm}{!}{\includegraphics{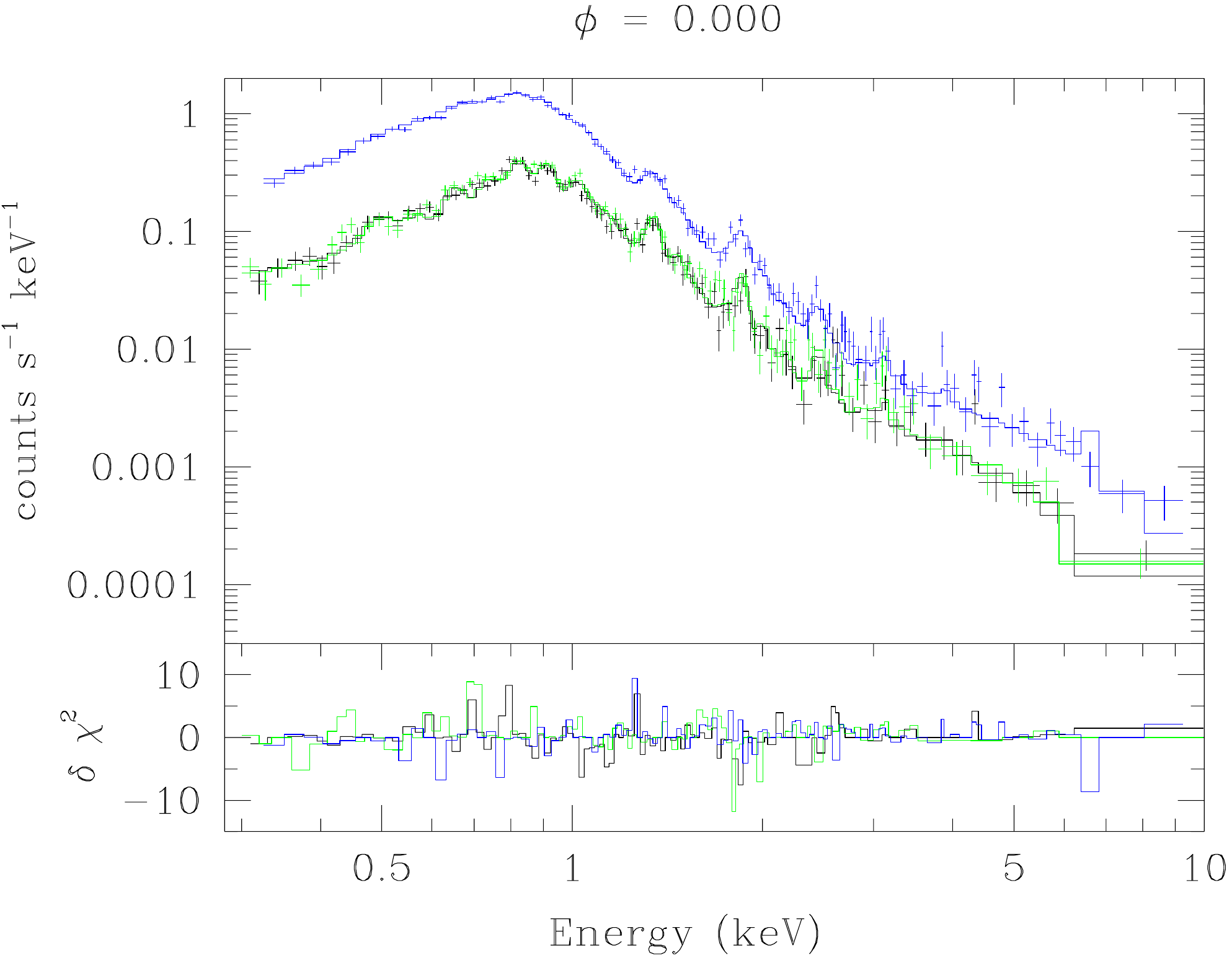}}
\end{minipage}
\begin{minipage}{8cm}
\resizebox{8cm}{!}{\includegraphics{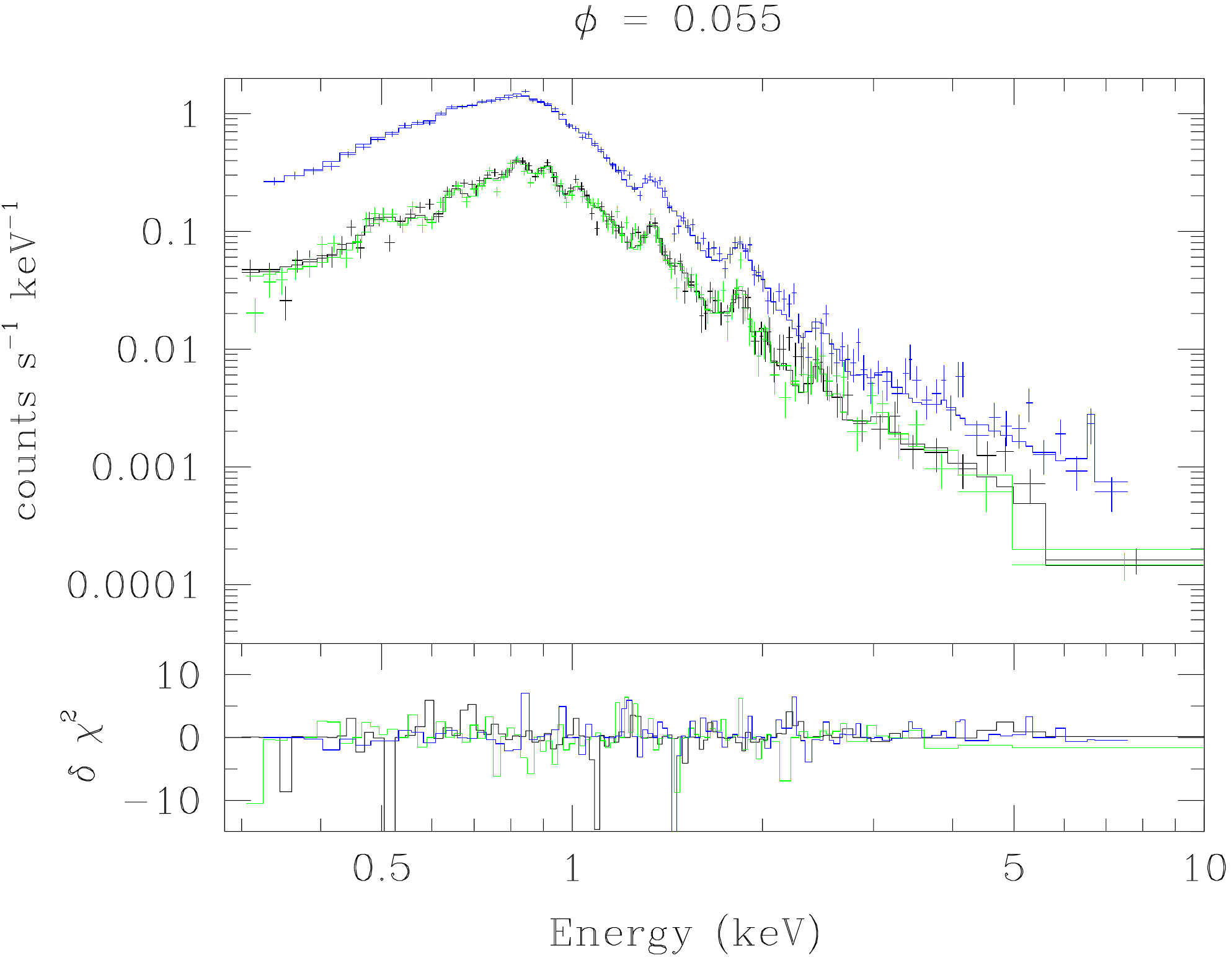}}
\end{minipage}
\caption{EPIC spectra of 9~Sgr as a function of orbital phase along with the best-fit thermal non-equilibrium (Eq.\,\ref{eq4}) models. The top panels illustrate the EPIC-pn (blue), EPIC-MOS1 (black) and EPIC-MOS2 (green) data. The bottom panels yield the contributions of the different energy bins to the $\chi^2$ of the fit with the sign of $\delta\chi^2$ being defined as the sign of the observation minus the model. All panels have the same vertical and horizontal scales. Note the relatively constant flux level in the soft part of the spectrum but also the strong changes in the level of the hard flux between $\phi = 0.627$ and the spectra taken close to periastron.\label{EPICplot}}
\end{figure*}
\begin{table*}[h!tb]
\caption{Observed fluxes as derived from the model fits in Sect.\,\ref{broad}.\label{Xflux}}
\begin{center}
\begin{tabular}{c c c c c}
\hline
\vspace*{-3mm}\\
$\phi$ & \multicolumn{4}{c}{$f_{\rm X}$ ($10^{-13}$\,erg\,cm$^{-2}$\,s$^{-1}$)} \\
\cline{2-5}
\vspace*{-3mm}\\
& (0.5 -- 1.0)\,keV & (1.0 -- 2.0)\,keV & (2.0 -- 10.0)\,keV & (0.5 -- 10.0)\,keV \\
\hline
0.627 & $11.95 \pm 0.26$ & $4.39 \pm 0.03$ & $0.62 \pm 0.12$ & $16.97 \pm 0.38$ \\
0.946 & $12.37 \pm 0.19$ & $5.16 \pm 0.05$ & $2.28 \pm 0.08$ & $19.80 \pm 0.22$ \\
0.000 & $13.00 \pm 0.20$ & $6.18 \pm 0.12$ & $2.69 \pm 0.11$ & $21.87 \pm 0.38$ \\
0.055 & $12.65 \pm 0.21$ & $5.42 \pm 0.07$ & $2.28 \pm 0.07$ & $20.33 \pm 0.26$ \\ 
\hline
\end{tabular}
\end{center}
\end{table*}

Unlike what is seen in some other long-period interacting wind O + O binaries (e.g.\ Cyg\,OB2 \#9, Naz\'e et al.\ \cite{YN12}), the high-energy tail does not reveal a strong Fe K line near 6.7\,keV. At phases $\phi = 0.946$ and $\phi = 0.055$, there are hints for the presence of such a line in the EPIC-pn spectra, which have the highest S/N ratio over this energy band, although only at a rather low level. No trace of this line is seen at $\phi = 0.627$, when the hard tail is very weak. What is more surprising is the lack of such an Fe K line in the spectrum taken at $\phi = 0.000$. We shall come back to this point in Sect.\,\ref{discussion}.\\

The {\tt vapec} models used above are based on the assumption of ionization equilibrium. Several authors (Usov \cite{Usov}, Zhekov \& Skinner \cite{ZS}, Zhekov \cite{Zhekov}, Pollock et al.\ \cite{Pollock}, amongst others) have drawn attention to the fact that the shocks in wide colliding wind binaries could be collisionless and that the conditions in the post-shock plasma could deviate from ionization equilibrium. This is motivated by the fact that the timescales needed to reach ionization equilibrium are longer than the dynamical timescales.

Zhekov (\cite{Zhekov}) proposed to adopt the quantity 
\begin{equation}
\Gamma_{\rm NEI} = 1.21\,\frac{\chi_e\,\dot{M}}{\overline{\mu}\,v_{\infty}^2\,d}
\end{equation}
as a criterion to estimate whether or not non-equilibrium ionization effects are important. In this relation, $\chi_e$ is the ratio between the electron number density and the nucleon number density, $\overline{\mu}$ is the mean atomic weight for nucleons and $d$ is the distance from the star to the shock in $10^{16}$\,cm. The mass-loss rate and the terminal wind velocity are expressed in units $10^{-5}$\,M$_{\odot}$\,yr$^{-1}$ and 1000\,km\,s$^{-1}$, respectively. According to Zhekov (\cite{Zhekov}), a value of $\Gamma_{\rm NEI} \gg 1$ indicates that such effects can be neglected, whereas they are important if $\Gamma_{\rm NEI} \leq 1$. To use this criterion, we first need to estimate the mass-loss rates and wind velocities of both components of 9~Sgr. For this purpose, we use the spectral types inferred in Paper II, along with typical effective temperatures, masses and luminosities from Martins et al.\ (\cite{Martins}). We then inject these numbers into the mass-loss recipe of Vink et al.\ (\cite{Vink}, their equation 24), assuming $v_{\infty}/v_{\rm esc} \simeq 2.6$ which leads to $v_{\infty} \simeq 2700$\,km\,s$^{-1}$, close to $v_{\infty} = 2750$\,km\,s$^{-1}$ as determined by Prinja et al.\ (\cite{Prinja}) on {\it IUE} spectra. This leads to estimates of the primary and secondary mass-loss rates of $\log{\frac{\dot{M}}{\rm M_{\odot}\,yr^{-1}}} = -5.46$ and $-5.91$ respectively. Using these mass-loss rates and terminal wind velocities, we estimate $\Gamma_{\rm NEI} \in [2.11\,\sin{i}, 12.21\,\sin{i}]$ for the primary and $\Gamma_{\rm NEI} \in [1.22\,\sin{i}, 7.07\,\sin{i}]$ for the secondary. The upper limit of the range of $\Gamma_{\rm NEI}$ corresponds to the periastron passage, whereas the lower one corresponds to apastron. If we were instead to adopt the stellar parameters found in Paper II\footnote{In Paper II, we used $\dot{M}_p = 9 \times 10^{-7}$\,M$_{\odot}$\,yr$^{-1}$, $v_{\infty, p} = 3500$\,km\,s$^{-1}$ and $\dot{M}_s = 5 \times 10^{-7}$\,M$_{\odot}$\,yr$^{-1}$, $v_{\infty, s} = 3100$\,km\,s$^{-1}$ for the primary and secondary stars, respectively. These parameters were based on CMFGEN (Hillier \& Miller \cite{CMFGEN}) models built to check consistency with the disentangled spectra. The $v_{\infty}$ values are close to $v_{\rm max}$ quoted by Prinja et al.\ (\cite{Prinja}).\label{note}}, then the results would be $\Gamma_{\rm NEI} \in [0.36\,\sin{i}, 2.09\,\sin{i}]$ for the primary and $\Gamma_{\rm NEI} \in [0.37\,\sin{i}, 2.12\,\sin{i}]$ for the secondary. We thus conclude that, non-equilibrium ionization effects are possibly present in the wind interaction zone of 9~Sgr, especially at phases away from periastron. Near periastron, our best estimate of $i$ (see Sect.\,\ref{orbital}) yields $\Gamma_{\rm NEI} \simeq 1.3$ for the stellar parameters from Paper II, suggesting also mild deviations from equilibrium at this orbital phase. 

To test the possibility of a non-equilibrium plasma, we have thus used the constant temperature plane parallel shock model {\tt vpshock} implemented under {\tt xspec}. This model was originally designed to study the thermal emission of supernova remnants (Borkowski et al.\ \cite{NEI}). The expression of our model was
\begin{eqnarray}
& & {\tt phabs}*[{\tt mtab\{wind_1\}*vapec(T_1)} +  \nonumber \\
& & {\tt mtab\{wind_2\}*(vapec(T_2) + vpshock(T_3)})] 
\label{eq4}
\end{eqnarray}
The parameters of the ISM absorption and of the first plasma component were again fixed to their values derived above. The resulting fit parameters are provided in Table\,\ref{EPICfits} and are illustrated in Fig.\,\ref{EPICplot}. There is a slight improvement in the fit quality compared to the models that assume ionization equilibrium, especially at phase 0.627. 
The {\tt vpshock} model features a parameter $\tau_u$, the ionization timescale of the non-equilibrium model, which in the original model of Borkowski et al.\ (\cite{NEI}) is defined as the product of the post-shock electron number density and the age of the supernova remnant. For 9~Sgr, the best fit $\tau_u$ values are roughly an order of magnitude lower than found by Pollock et al.\ (\cite{Pollock}) in the case of the archetypal colliding wind system WR~140 (WC7pd + O5.5\,fc, P = 7.93\,yrs, $e = 0.89$), but so are the wind densities in 9~Sgr compared to those of WR~140.

In the strong shock limit, the kinetic energy normal to the shock is converted into heat according to (e.g.\ Stevens et al.\ \cite{SBP})
\begin{equation}
k\,T_s = \frac{3}{16}\,m_p\,v_{\perp}^2
\label{eq5}
\end{equation}
where $T_s$, $m_p$ and $v_{\perp}$ are the post-shock temperature, the particle mass and the particle's pre-shock velocity perpendicular to the shock, respectively. If we assume that the winds have essentially solar composition and collide with $v_{\perp} = v_{\infty} = 2750$\,km\,s$^{-1}$, we can estimate that, for an average particle mass of $10^{-24}$\,g, the shocked plasma would be heated to about 8.9\,keV near the binary axis where the collision is head-on. Away from this axis, the temperature of the plasma should be lower as the shocks become more oblique. Because of the large difference in mass between the electrons and the ions, the immediate post-shock electron temperature should be much lower than the temperature of the ions. Equalization of the temperatures should then occur via Coulomb interactions between the electrons and the ions. However, because of the low plasma density, this process is rather slow. According to Usov (\cite{Usov}), equalization of the ion and electron temperatures occurs on a timescale 
\begin{equation}
t_{\rm eq} = 10\,\left(\frac{T_e}{10^7\,{\rm K}}\right)^{3/2}\,\left(\frac{n_e}{10^{10}\,{\rm cm}^{-3}}\right)^{-1}\,s
\label{eq6}
\end{equation}
With the various estimates of the mass-loss rates and wind velocities outlined above, and considering that the electron temperature is reflected by the value of $kT_3 \simeq 6$\,keV derived for the {\tt vpshock} models in Table\,\ref{EPICfits}, the equalization timescale given by Eq.\,\ref{eq6} would be of order $10^5$ to $10^7$\,s depending on the orbital phase. Since this timescale is long compared to the flowing timescale of the plasma, the shocks are collisionless (e.g.\ Zhekov \& Skinner \cite{ZS}, Zhekov \cite{Zhekov}, Pollock et al.\ \cite{Pollock}). If we assume instead that the relevant electron temperature in Eq.\,\ref{eq6} is the immediate post-shock temperature given by Eq.\,\ref{eq5}, the timescale  would be considerably reduced by a factor $\sim 10^{-5}$. Therefore, the initial steps of the temperature equalization proceed rather quickly and the actual electron temperature is likely to be intermediate between the immediate post-shock value estimated from Eq.\,\ref{eq5} and the ion temperature. From an observational point of view, the results of our spectral fits suggest that $T_e$ is actually rather close to the average particle temperature given by equation\,\ref{eq6}.\\ 

Since the X-ray spectrum of 9~Sgr reveals only a very weak Fe K line, one could also wonder whether the hard tail of the spectrum is fully thermal or actually partially non-thermal. Indeed, the observed synchrotron radio emission implies the existence of a population of relativistic electrons that could produce a non-thermal X-ray emission through inverse Compton scattering. Therefore, we have also considered the possibility of a non-thermal power-law model contributing to the hard tail of the X-ray spectrum. For this purpose we used a model of the kind
\begin{eqnarray}
& & {\tt phabs}*[{\tt mtab\{wind_1\}*vapec(T_1)} +  \nonumber \\
& & {\tt mtab\{wind_2\}*(vapec(T_2) + power})] 
\label{eq3}
\end{eqnarray}
with the parameters of the first plasma component again fixed to the above values. The resulting fits (see lower part of Table\,\ref{EPICfits}) are of very similar quality to those obtained with the purely thermal models with ionization equilibrium. Except for the $\phi = 0.627$ observation, the best-fit photon index of the power-law is close to $2.5$. We will come back to this point in Sect.\,\ref{discussion}.\\ 

The results of this section suggest that the non-equilibrium ionization models produce the best fits. However, the model parameters are not to be taken too literally and the fits should rather be regarded as a convenient way to represent the X-ray spectral energy distribution. For instance, we note that there exists a degeneracy between the wind column density of the medium and hard component and their normalization. 

\begin{figure}
\resizebox{9cm}{!}{\includegraphics{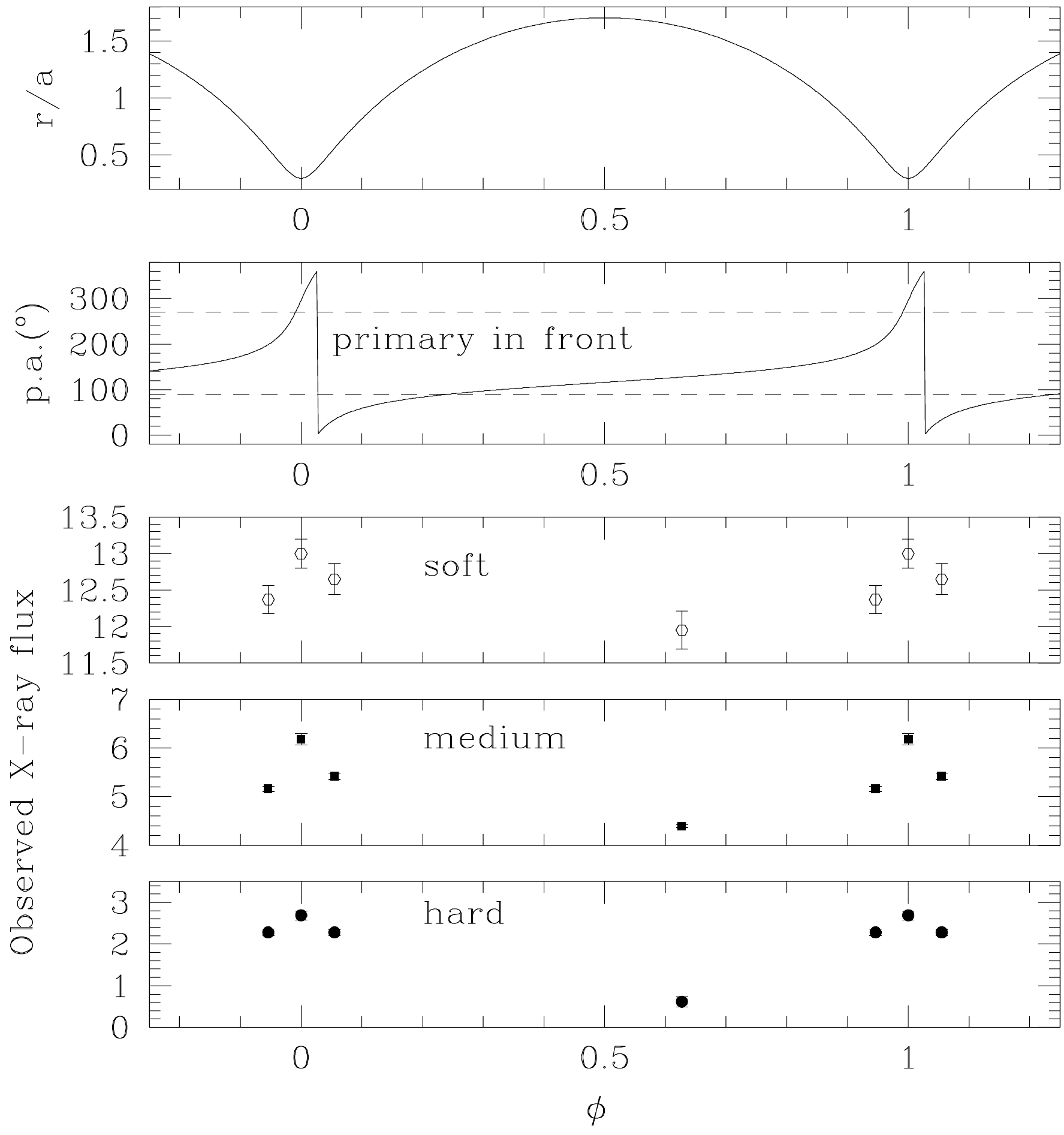}}
\caption{The top two panels illustrate the variations of the orbital separation ($r/a$) and the position angle (defined as p.a. = 0$^{\circ}$ at the conjunction with the secondary star in front) as a function of orbital phase. The three panels at the bottom yield the variations of the observed fluxes in units $10^{-13}$\,erg\,cm$^{-2}$\,s$^{-1}$ over the three energy bands (soft: 0.5 -- 1.0\,keV; medium: 1.0 -- 2.0\,keV; hard: 2.0 -- 10.0\,keV). \label{Xvariations}}
\end{figure}

The various fits performed in this section allow us to derive the observed fluxes and to estimate their uncertainties from the dispersion of the values obtained with the different model assumptions. They are given in Table\,\ref{Xflux} for different energy bands. The variations of these fluxes as a function of phase are shown in Fig.\,\ref{Xvariations}.

It appears clearly that the flux is largest in all three energy bands at periastron. However, the relative amplitude of the variations is quite different, with only about 8\% variations in the soft (0.5 -- 1.0\,keV) band, about 1/3 of the flux varying in the medium (1.0 -- 2.0\,keV) band and a factor of 4 variations in the hard (2.0 -- 10.0\,keV) energy domain. This is further evidence that the bulk of the soft X-ray flux is not arising in the wind interaction zone, whereas most of the hard X-ray flux probably comes from this interaction zone. 

\section{Discussion \label{discussion}}
In an interacting wind binary system, the binary period influences the nature of the interaction. In close massive binary systems with short orbital periods, the gas density in the interaction region is sufficiently high for radiative cooling to become very efficient and these systems are thus in the radiative regime (Stevens et al.\ \cite{SBP}). Conversely, in wide long-period binaries the postshock gas is less dense, and as a result these wind collisions are in an adiabatic regime and cool very slowly (Stevens et al.\ \cite{SBP}). 

According to Stevens et al.\ (\cite{SBP}), the nature of the wind interaction can be estimated from the ratio $\chi$ between the cooling time and the escape time. Using the wind parameters estimated by means of the Vink et al.\ (\cite{Vink}) formalism and adopting the orbital separation at periastron passage, equation 8 of Stevens et al.\ (\cite{SBP}) yields $\chi = 64/\sin{i}$ and $107/\sin{i}$ for the shocked primary and secondary winds, respectively. If we instead use the parameters from Paper II (see footnote\,\ref{note}), we obtain $\chi = 601/\sin{i}$ and $467/\sin{i}$, reinforcing our conclusion about the shocks being adiabatic. Therefore, we find that, within a large margin on the stellar parameters, both winds are expected to be fully adiabatic all over the orbit. For the Vink et al.\ (\cite{Vink}) wind parameters, the square root of the wind momentum ratio ${\cal R} = \left(\frac{\dot{M}_p\,v_{\infty,p}}{\dot{M}_s\,v_{\infty,s}}\right)^{1/2}$ amounts to 1.68 and from the formalism of Cant\'o et al.\ (\cite{Canto}), we can then estimate a shock half opening angle of $71.4^{\circ}$ (see Fig.\,\ref{ellipse}). 

Still, using the formalism of Cant\'o et al.\ (\cite{Canto}) along with the above stellar wind parameters, we can evaluate the emission measure $\int n_e\,n_H\,dV$ of the shocked plasma. In this way, we obtain a value of $3.4 \times 10^{56}$\,cm$^{-3}$ at periastron passage. This number can be compared to the normalization of the hot plasma component ($kT_3$) in the thermal non-equilibrium ionization models of Table\,\ref{EPICfits}. Adopting a distance of 1.8\,kpc, the normalization parameter at $\phi = 0.000$ yields an emission measure of $7.3 \times 10^{54}$\,cm$^{-3}$, i.e.\ about a factor 46 lower than the estimate based on the Cant\'o et al.\ (\cite{Canto}) formalism. This discrepancy is actually not as severe as it looks at first sight as part of it could stem from the adopted mass-loss rates. Indeed, over recent years, empirical mass-loss rate determinations, accounting for small-scale clumping, suggest that the recipe of Vink et al.\ (\cite{Vink}) leads to overpredictions of the mass-loss by a factor $\sim 3$ (Bouret et al.\ \cite{Bouret}). For instance, adopting the wind parameters from Paper II (see footnote\,\ref{note}) reduces the discrepancy to a factor 3.5.

For adiabatic wind interaction regions in wide eccentric binary systems, the X-ray luminosity from the wind-wind collision is expected to vary as $1/r$ where $r$ is the orbital separation (Stevens et al.\ \cite{SBP}). A spectacular illustration of such a $1/r$ X-ray flux variation is the O5-5.5\,I + O3-4\,III system Cyg\,OB2 \#9 which has a period of 2.36\,yr and an eccentricity of $e = 0.7$ (Naz\'e et al.\ \cite{YN12}). 

\begin{figure}
\resizebox{9cm}{!}{\includegraphics{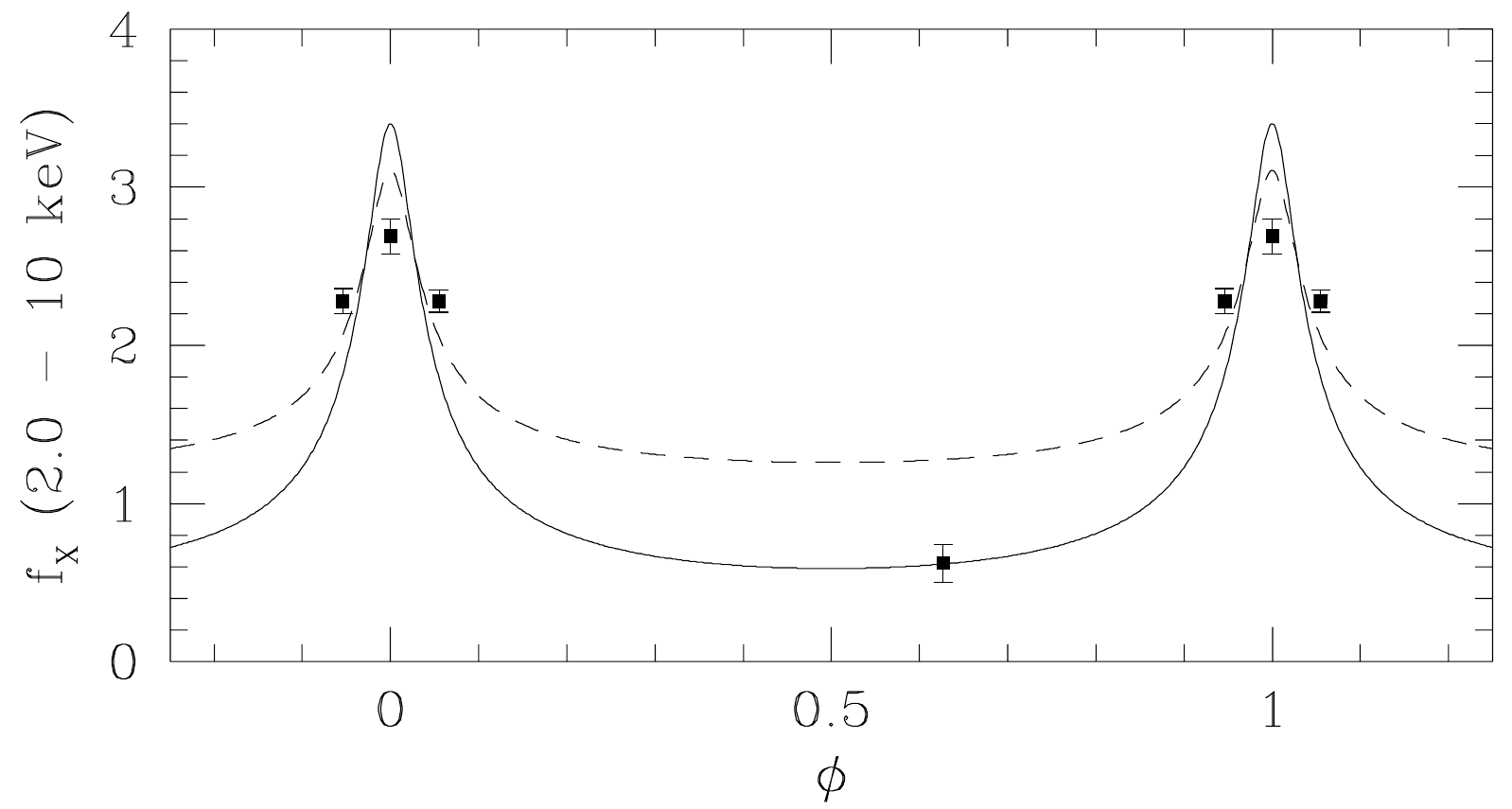}}
\caption{Observed X-ray flux of 9~Sgr in the hard band. The solid line yields a $1/r$ relation scaled to match the flux level at phase $\phi = 0.627$, whilst the dashed line corresponds to the least-square fit to all four data points of a $1/r$ relation including an offset to account for emission not coming from the wind interaction zone (see text).\label{X1od}}
\end{figure}

For 9~Sgr, the results in the present paper show indeed that the X-ray emission is maximum at periastron passage. However, as we have shown in Sects.\,\ref{highres} and \ref{broad}, the bulk of the soft and medium emission does not come from the wind-wind interaction, but rather from the intrinsic X-ray emission of both stars. On the other hand, the flux in the hard band undergoes a modulation which, at first sight, is quite close to the expected $1/r$ variation (see Fig.\,\ref{X1od}), suggesting that this part of the emission traces the wind interaction zone. 

If we scale the $1/r$ relation in such a way to match the level of the hard emission at phase $0.627$, i.e.\ assuming that all the hard flux at this phase arises in the wind interaction zone, we find that the observed flux level at periastron falls short by 25\% compared to the expected emission level (see Fig.\,\ref{X1od}). Alternatively, if we perform a least-square fit of a relation of the kind $f_{\rm X} (2.0 - 10) = {\rm coeff}_1*(1/r) + {\rm coeff}_2$, where ${\rm coeff}_2$ accounts for a constant emission in the hard band not due to colliding winds, we obtain the dashed line in Fig.\,\ref{X1od}. This fitting relation improves the agreement around periastron passage, but the data point at $\phi = 0.627$ falls well below the expected level. It thus seems that the hard X-ray flux of 9~Sgr follows a somewhat more complex behaviour than the simple $1/r$ relation. 

Based on hydrodynamical simulations, Pittard \& Parkin (\cite{PP}) indeed predicted the existence of deviations from this behaviour. However, their simulations refer to binary systems with much shorter orbital periods than 9~Sgr. In these short-period systems, the impact of orbital motion on the wind interaction zone is important and the shocks may change from being adiabatic to partially or even fully radiative at periastron. In 9~Sgr, none of these effects is expected to play a role. 
Radiative braking (Gayley et al.\ \cite{Gayley}) also cannot explain the deviations from the $1/r$ relation. Indeed, radiative braking is most relevant for systems with separations up to 100\,R$_{\odot}$ and large contrasts in wind strengths (Gayley et al.\ \cite{Gayley}). Using the stellar and orbital parameters of the components of 9~Sgr, we have computed the location of 9~Sgr in the homology diagram of Gayley et al.\ (\cite{Gayley}, their Fig.\,5). We estimated a scaled momentum ratio of 2.6 and a scaled separation of 47.2 at periastron passage. Hence, the system falls well outside the area of the parameter space concerned by radiative braking.
A more promising avenue could be radiative inhibition (Stevens \& Pollock \cite{SP}), although this effect also strongly decreases with orbital separation (see Fig.\,5 of Stevens \& Pollock \cite{SP}). In their 3-D hydrodynamical simulations of Cyg\,OB2 \#9, Parkin et al.\ (\cite{Parkin}) indeed reported a drop of the pre-shock wind velocity by $\simeq 18$\% at periastron. 9~Sgr and Cyg\,OB \#9 have very similar eccentricities, but 9~Sgr has a two times wider orbital separation and its components are slightly less luminous than those of Cyg\,OB2 \#9. Both effects should reduce the importance of radiative inhibition in 9~Sgr compared to Cyg\,OB2 \#9. A detailed assessment of the importance of radiative inhibition requires 3-D hydrodynamic simulations as performed by Parkin et al.\ (\cite{Parkin}) which is beyond the scope of the present work.\\ 

\begin{figure}
\resizebox{9cm}{!}{\includegraphics{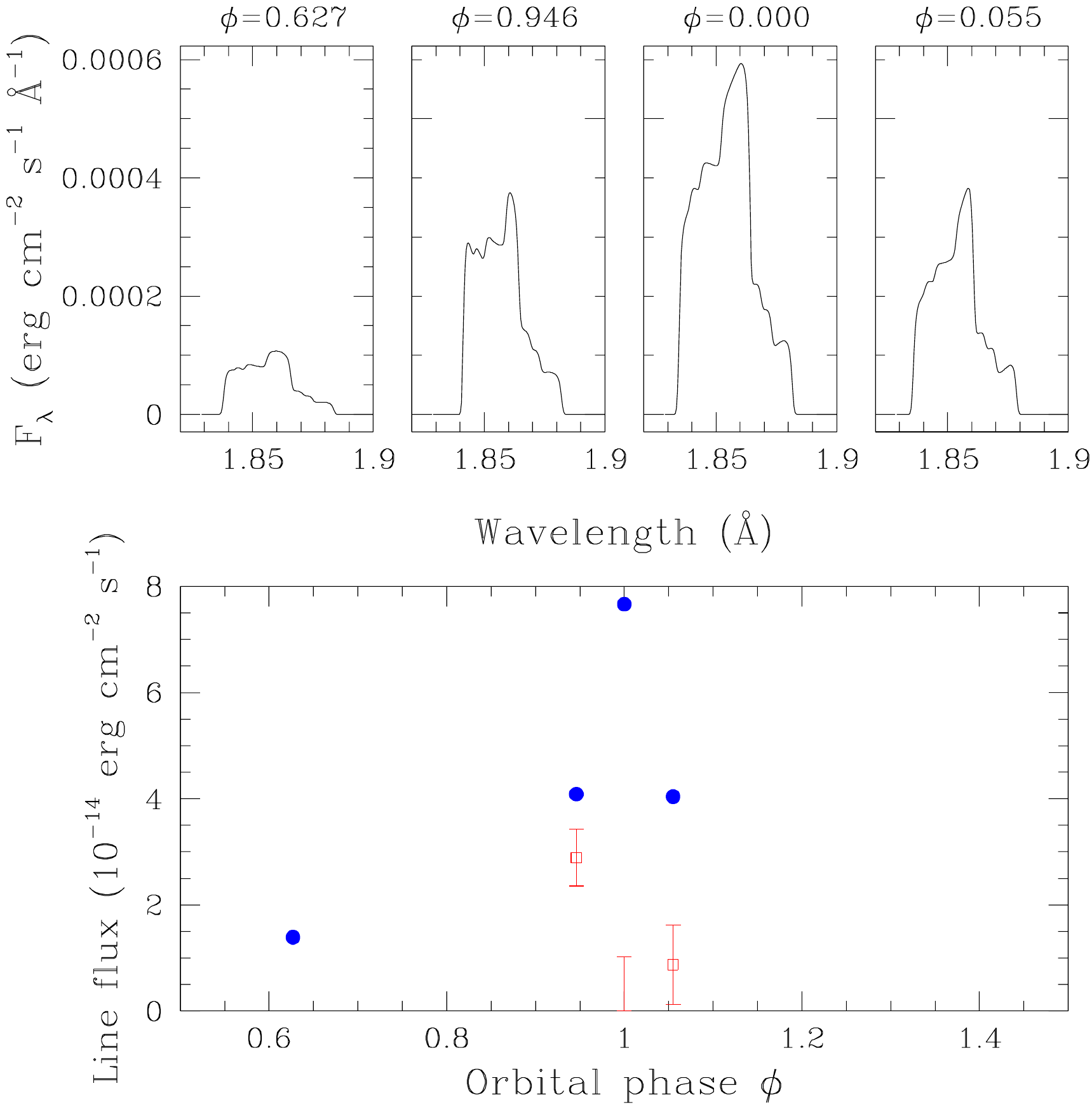}}
\caption{Top row: synthetic profiles of the Fe\,{\sc xxv} line blend as predicted by the model of Rauw et al.\ (\cite{RM}, see also text). Bottom: variation of the integrated fluxes of the synthetic line profiles (filled symbols) as a function of orbital phase. The synthetic fluxes were scaled to match the observed emission measure of the hard plasma component at $\phi = 0.000$. The measured fluxes at $\phi = 0.946$ and $0.055$ are shown by the open symbols. An upper limit on the flux at $\phi = 0.000$ is also shown.\label{profile}}
\end{figure}

Another puzzle is the behaviour of the Fe\,{\sc xxv} blend at 6.7\,keV. As becomes clear from Fig.\,\ref{EPICplot}, the Fe K line is seen in the EPIC-pn data at phase $\phi = 0.946$, and to a lesser extent in the data from $\phi = 0.055$. However, no such line is detected at $\phi = 0.627$ or $\phi = 0.000$. Whilst the lack of detection at $\phi = 0.627$ could be attributed to the low level of the hard tail at this phase, the result at $\phi = 0.000$ is rather unexpected. To quantify what one would expect to observe, we have used the model of Rauw et al.\ (\cite{RM}) to compute synthetic line profiles from the wind interaction region. This model uses the formalism of Cant\'o et al.\ (\cite{Canto}) to compute the shape of the contact discontinuity and the surface density of the shocked material. The winds are assumed to collide at $v_{\infty}$. The post-shock densities and temperatures are computed from the Rankine-Hugoniot conditions for a strong adiabatic shock assuming ionization equilibrium and equal electron and ion temperatures. The Fe K line emissivity is evaluated at the temperature of the post-shock plasma. The material in the wind interaction zone is assumed to flow with a velocity tangent to the contact discontinuity. This velocity is then projected onto the line-of-sight towards the observer to compute the line profile as a function of wavelength. The stellar wind parameters were adopted from Paper II (see footnote\,\ref{note}) since they were found to provide a reasonably good agreement with the observed emission measure. We further assumed an orbital inclination of $45^{\circ}$. The resulting synthetic profiles of the Fe\,{\sc xxv} blend are shown in the upper panel of Fig.\,\ref{profile} at the four orbital phases observed with {\it XMM-Newton}. Whilst our observations lack of course the spectral resolution required to compare with these predictions, we can nevertheless compare the integrated fluxes. It becomes immediately clear that the model predicts the strongest line at periastron. In fact, since we assume an adiabatic wind interaction zone in our model, the theoretical line flux varies as $1/r$. 

In the bottom panel of Fig.\,\ref{profile}, we compare the observed fluxes of the Fe K line (determined by fitting a power-law + Gaussian model to the data above 2.0\,keV\footnote{Note that no Fe K line was detected at phases 0.627 and 0.000. For the latter phase we nevertheless obtained an upper limit on the flux.}) to the theoretical values obtained by integrating the synthetic profiles. The fluxes of the synthetic profiles were further scaled to match the observed emission measure of the hard component in the non-equilibrium thermal models at $\phi = 0.000$ (see Table\,\ref{EPICfits}). As we can see, the Fe K line flux clearly does not follow the $1/r$ trend. Whilst the model reasonably matches the line flux at phase $\phi=0.946$, there is a huge discrepancy at periastron and a severe discrepancy at $\phi = 0.055$. A line as strong as predicted by our model should have been easily detected in our observations. This clearly hints at a change in the properties of the post-shock plasma during periastron passage.\\  

One possibility to explain the oddities of the high-energy tail would be that this emission might, at least partially, be affected by non-thermal processes. As indicated by the presence of a synchrotron radio emission, the winds of 9~Sgr host a population of relativistic electrons. Together with the photospheric UV radiation, these relativistic electrons could produce a non-thermal X-ray emission. Bell (\cite{Bell}) showed that a population of relativistic electrons accelerated by shocks should have a power-law energy distribution with an index $n = (\xi +2)/(\xi -1)$ where $\xi$ is the compression ratio of the shock. Inverse Compton scattering by these electrons then leads to a power-law spectrum with a photon index $\Gamma = (n+1)/2$ (Chen \& White \cite{CW}). Since we have shown above that the shock between the stellar winds of 9~Sgr should be adiabatic, we find that $\xi = 4$, leading to $\Gamma = 1.5$. Our best-fit partially non-thermal models in Table\,\ref{EPICfits} yield much steeper power-laws, although this result must be considered with some caution as the photon index of the power-law could be influenced by degeneracies between the various components of model Eq.\,\ref{eq3}. Hence, the broad-band spectral fits provide no unambiguous evidence for the detection of such a non-thermal X-ray emission in the {\it XMM-Newton} energy band.

If such a non-thermal emission nevertheless exists, the flux due to inverse Compton scattering should be directly proportional to the density of the seed UV radiation field in the wind interaction zone which scales as $1/r^2$. As we have shown in Sect.\,\ref{broad}, the observed flux shows a shallower trend than $1/r$ and is thus certainly not compatible with $1/r^2$. However, here we implicitly assume that the population of relativistic electrons does not change with orbital phase. A different approach to the question would be to use the simple formula 
\begin{equation}
L_{\rm IC} = \frac{U_{\rm ph}}{U_{\rm B}}\,L_{\rm Sync}
\label{ICsync}
\end{equation}
(e.g.\ Pittard \& Dougherty \cite{PD}) where $L_{\rm IC}$ and $L_{\rm Sync}$ are the inverse Compton and synchrotron luminosities, respectively, whilst $U_{\rm ph}$ and $U_{\rm B}$ stand for the seed photon and the magnetic field energy densities. The latter of these quantities depends on the strength of the magnetic field in the wind interaction zone. This quantity is currently not known. Hubrig et al.\ (\cite{Hubrig0,Hubrig1}) claimed the detection of a magnetic field in 9~Sgr with longitudinal field strengths varying between $-265$ and $+242$\,G due to rotational modulation. However, Neiner et al.\ (\cite{Neiner}) detected no magnetic field in a sample of particle-accelerating colliding wind binaries. For 9~Sgr, Neiner et al.\ (\cite{Neiner}) reported upper limits on the longitudinal magnetic field of $|B_l| < 64$\,G, resulting in an upper limit on the dipolar field strength of $B_{\rm pol, max} = 605$\,G. Either the magnetic field involved in the production of the synchrotron emission is stellar but sufficiently weak to remain below the detection threshold (e.g.\ Parkin et al.\ \cite{Parkin}), or it is generated inside the wind interaction zone. In either case, it is very challenging to estimate $U_{\rm B}$. 

Still, a comparison between the variations of the hard X-ray flux from the wind interaction zone and the 2\,cm radio lightcurve of Paper III (Blomme \& Volpi \cite{BV}) reveals an interesting result. Indeed, the best-fit model for the radio data of Paper III predicts a variation of the 2\,cm radio flux by approximately a factor 4 between apastron and periastron\footnote{In 9~Sgr, the colliding wind region seems to be outside the radio photosphere (at $\lambda = 2$\,cm) at all orbital phases and the variations of the observed radio flux should thus not be affected by absorption by the unshocked winds.}. This is the same ratio as we observe for the variation of the hard X-ray flux (see Fig.\,\ref{X1od}), suggesting a connection between the particle acceleration and the level of thermal X-ray emission.  

Pittard \& Dougherty (\cite{PD}) studied such a connection in the case of WR~140 notably emphasizing the role of modified shocks. Shock modification arises from the diffusion of non-thermal ions upstream of the shock. These non-thermal ions create a so-called shock precursor where the back pressure exerted by the non-thermal ions on the inflowing gas gradually reduces the pre-shock velocity of the latter. This leads to a softer X-ray spectrum and might actually help equilibrating the electron and ion temperatures. As pointed out by Pittard \& Dougherty (\cite{PD}), the degree of shock modification could be dependent on the orbital phase. 

We thus suggest that the deviation of the near periastron hard X-ray flux from the $1/r$ relation stems from the fact that a substantial fraction of the energy is placed into accelerated particles with an efficiency that depends on the orbital phase. How such a process affects the Fe {\sc xxv} line emission remains to be seen. If this scenario is correct, an obvious question that comes to mind is why we observe large deviations from the $1/r$ relation in 9~Sgr but not in Cyg\,OB2 \#9 (Naz\'e et al.\ \cite{YN12}) which is also a prominent non-thermal radio emitter. A detailed modelling of these systems as done by Pittard \& Dougherty (\cite{PD}) for WR~140 would certainly shed new light onto the problem, but such a study is beyond the scope of the present paper.

\section{Conclusions}
We have presented the results of optical and X-ray spectroscopy of 9~Sgr gathered around its most recent periastron passage. Our optical spectra have enabled us to refine the orbital solution and revise the orbital period to 9.1\,yrs. With the new ephemerides at hand, we have interpreted the four X-ray observations of this system. Whilst the soft emission is very likely dominated by the intrinsic emission from the stars, as indicated by the lack of a substantial phase-dependence, on the one hand, and the properties of the high-resolution X-ray spectrum, on the other hand, the hard emission undergoes a clear modulation with orbital phase. The X-ray flux is largest at periastron as expected for an adiabatic wind interaction zone. However, the variations deviate from the expected $1/r$ scaling relation, and we suggest that this situation could reflect the impact of radiative inhibition as well as of particle acceleration on the shock properties. The forthcoming analysis of the radio data (Blomme et al.\ in prep.) will provide a first check of the plausibility of this scenario. 
 
\acknowledgement{We are grateful to the observers at the Mercator Telescope for collecting the data of our HERMES service mode programme. We thank Dr.\ H.\ Sana for sharing preliminary results of his interferometry campaign with us. The Li\`ege team acknowledges support through an ARC grant for Concerted Research Actions, financed by the Federation Wallonia-Brussels, from the Fonds de la Recherche Scientifique (FRS/FNRS), as well as through an XMM PRODEX contract.}


\begin{thebibliography}{}
\bibitem[1984]{ABC} Abbott, D.C., Bieging, J.H., \& Churchwell, E.\ 1984, ApJ, 280, 671
\bibitem[1986]{ABCT} Abbott, D.C., Bieging, J.H., Churchwell, E., \& Torres, A.V.\ 1986, ApJ, 303, 239
\bibitem[2015]{Aldoretta} Aldoretta, E.J., Caballero-Nieves, S.M., Gies, D.R., et al.\ 2015, AJ, 149, 26
\bibitem[1989]{AG} Anders, E., \& Grevesse, N.\ 1989, Geochimica et Cosmochimica Acta, 53, 197
\bibitem[1996]{xspec} Arnaud, K.A.\ 1996, in Astronomical Data Analysis Software and Systems V, eds.\ G.\ Jacoby, \& J.\ Barnes, ASP Conf.\ Series, 101, 17
\bibitem[1992]{BC}Ba\l uci\'nska-Church, M., \& McCammon, D.\ 1992, ApJ, 400, 699
\bibitem[1996]{Baranne} Baranne, A., Queloz, D., Mayor, M., et al.\ 1996, A\&AS, 119, 373
\bibitem[1978]{Bell} Bell, A.R.\ 1978, MNRAS, 182, 147
\bibitem[2014]{BV} Blomme, R., \& Volpi, D.\ 2014, A\&A, 561, A18, (Paper III)
\bibitem[1972]{Blumenthal} Blumenthal, G.R., Drake, G.W.F., \& Tucker, W.H.\ 1972, ApJ, 172, 205
\bibitem[1978]{Bohlin} Bohlin, R.C., Savage, B.D., \& Drake, J.F.\ 1978, ApJ, 224, 132
\bibitem[2012]{Bouret} Bouret, J.-C., Hillier, D.J., Lanz, T., \& Fullerton, A.W.\ 2012, A\&A, 544, A67
\bibitem[2001]{NEI} Borkowski, K.J., Lyerly, W.J., \& Reynolds, S.P.\ 2001, ApJ, 548, 820
\bibitem[1996]{Canto} Cant\'o, J., Raga, A.C., \& Wilkin, F.P.\ 1996, ApJ, 469, 729
\bibitem[1991]{CW} Chen, W., \& White, R.L.\ 1991, ApJ, 366, 512
\bibitem[1976]{Chere} Cherepashchuk, A.M.\ 1976, SvAL, 2, 138
\bibitem[1991]{CG} Chlebowski, T., \& Garmany, C.D.\ 1991, ApJ, 368, 241
\bibitem[1996]{Corcoran} Corcoran, M.F.\ 1996, RevMexAA Conference Series, 5, 54
\bibitem[2004]{Damiani} Damiani, F., Flaccomio, E., Micela, G., Sciortino, S., Harnden, F.R.Jr., \& Murray, S.S.\ 2004, ApJ, 608, 781
\bibitem[2001]{RGS} den Herder, J.W., Brinkman, A.C., Kahn, S.M., et al.\ 2001, A\&A, 365, L7
\bibitem[1994]{DS} Diplas, A., \& Savage, B.D.\ 1994, ApJS, 93, 211
\bibitem[2000]{DW} Dougherty, S.M., \& Williams, P.M.\ 2000, MNRAS, 319, 1005
\bibitem[1990]{Alex} Fullerton, A.W.\ 1990, PhD thesis, University of Toronto
\bibitem[1969]{GJ} Gabriel, A.H., \& Jordan, C.\ 1969, MNRAS, 145, 241
\bibitem[1997]{Gayley} Gayley, K.G., Owocki, S.P., \& Cranmer, S.R.\ 1997, ApJ, 475, 786
\bibitem[2001]{Gosset} Gosset, E., Royer, P., Rauw, G., Manfroid, J., \& Vreux, J.-M.\ 2001, MNRAS, 327, 435
\bibitem[2012]{Gudennavar} Gudennavar, S.B., Bubbly, S.G., Preethi, K., \& Murthy, J.\ 2012, ApJS, 199, 8
\bibitem[1985]{HMM} Heck, A., Manfroid, J., \& Mersch, G.\ 1985, A\&AS, 59, 63
\bibitem[2013]{zetaPup} Herv\'e, A., Rauw, G., \& Naz\'e, Y.\ 2013, A\&A, 551, A83
\bibitem[1998]{CMFGEN} Hillier, D.J., \& Miller, D.L.\ 1998, ApJ, 496, 407
\bibitem[2008]{Hubrig0} Hubrig, S., Sch\"oller, M., Schnerr, R.S., Gonz\'alez, J.F., Ignace, R., \& Henrichs, H.F.\ 2008, A\&A, 490, 793
\bibitem[2013]{Hubrig1} Hubrig, S., Sch\"oller, M., Ilyin, I., et al.\ 2013, A\&A, 551, A33
\bibitem[2001]{Jansen} Jansen, F., Lumb, D., Altieri, B., et al.\ 2001, A\&A, 365, L1
\bibitem[1999]{Kaufer} Kaufer, A., Stahl, O., Tubbesing, S., et al.\ 1999, The Messenger, 95, 8 
\bibitem[1965]{LK} Lafler, J., \& Kinman, T.D.\ 1965, ApJS, 11, 216
\bibitem[2005]{Martins} Martins, F., Schaerer, D., \& Hillier, D.J.\ 2005, A\&A, 436, 1049
\bibitem[2004]{HD108} Naz\'e, Y., Rauw, G., Vreux, J.-M., \& De Becker, M.\ 2004, A\&A, 417, 667
\bibitem[2009]{Naze} Naz\'e, Y.\ 2009, A\&A, 506, 1055
\bibitem[2008]{YN08} Naz\'e, Y., De Becker, M., Rauw, G., \& Barbieri, C.\ 2008, A\&A, 483, 543
\bibitem[2010]{YN10} Naz\'e, Y., Damerdji, Y., Rauw, G., et al.\ 2010, ApJ, 719, 634
\bibitem[2011]{CarinaChandra} Naz\'e, Y., Broos, P.S., Oskinova, L., et al.\ 2011, ApJS, 194, 7
\bibitem[2012a]{YN12b} Naz\'e, Y., Flores, A.C., \& Rauw, G.\ 2012a, A\&A, 538, A22
\bibitem[2012b]{YN12} Naz\'e, Y., Mahy, L., Damerdji, Y., et al.\ 2012b, A\&A, 546, A37
\bibitem[2013]{zetaPup2} Naz\'e, Y., Oskinova, L.M., \& Gosset, E.\ 2013, ApJ, 763, 143 
\bibitem[2015]{Neiner} Neiner, C., Grunhut, J., Leroy, B., De Becker, M., \& Rauw, G.\ 2015, A\&A, 575, A66
\bibitem[2014]{Parkin} Parkin, E.R., Pittard, J.M., Naz\'e, Y., \& Blomme, R.\ 2014, A\&A, 570, A10
\bibitem[2009]{Pittard} Pittard, J.M.\ 2009, MNRAS, 396, 1743
\bibitem[2006]{PD} Pittard, J.M., \& Dougherty, S.M.\ 2006, MNRAS, 372, 801
\bibitem[2010]{PP} Pittard, J.M., \& Parkin, E.R.\ 2010, MNRAS, 403, 1657
\bibitem[1987]{Pollock1} Pollock, A.M.T.\ 1987, ApJ, 320, 283
\bibitem[2005]{Pollock} Pollock, A.M.T., Corcoran, M.F., Stevens, I.R., \& Williams, P.M.\ 2005, ApJ, 629, 482
\bibitem[2001]{Porquet} Porquet, D., Mewe, R., Dubau, J., Raassen, A.J.J., \& Kaastra, J.S.\ 2001, A\&A, 376, 1113
\bibitem[1976]{PU} Prilutskii, O.F., \& Usov, V.V.\ 1976, SvA, 20, 2
\bibitem[1990]{Prinja} Prinja, R.K., Barlow, M.J., \& Howarth, I.D.\ 1990, ApJ, 361, 607
\bibitem[2011]{Raskin} Raskin, G., van Winckel, H., Hensberge, H., et al.\ 2011, A\&A, 526, A69
\bibitem[2013]{Leuven} Rauw, G.\ 2013, in Setting a New Standard in the Analysis of Binary Stars, eds.\ K.\ Pavlovski, A.\ Tkachenko \& G.\ Torres, EAS Publication Series, 64, 59 
\bibitem[2002a]{PaperI} Rauw, G., Blomme, R., Waldron, W.L., et al.\ 2002a, A\&A, 394, 993, (Paper I)
\bibitem[2002b]{NGC6530} Rauw, G., Naz\'e, Y., Gosset, E., et al.\ 2002b, A\&A, 395, 499
\bibitem[2005]{jenam} Rauw, G., Sana, H., Gosset, E., et al.\ 2005, in Massive Stars and High Energy Emission in OB Associations, Proc. JENAM 2005, Distant Worlds, eds.\ G. Rauw, Y. Naz\'e, R. Blomme, E. Gosset, 85, http://www.astro.ulg.ac.be/RPub/Colloques/JENAM/proceedings
\bibitem[2012]{Paper2} Rauw, G., Sana, H., Spano, M., Gosset, E., Mahy, L., De Becker, M., \& Eenens, P.\ 2012, A\&A, 542, A95, (Paper II)
\bibitem[2015]{CygOB2Chandra} Rauw, G., Naz\'e, Y., Wright, N.J., et al.\ 2015, ApJS, 221, 1
\bibitem[2015]{lamCep} Rauw, G., Herv\'e, A., Naz\'e, Y., et al.\ 2015, A\&A, 580, A59
\bibitem[2016a]{RM} Rauw, G., Mossoux, E., \& Naz\'e, Y.\  2016a, New Astronomy, 43, 70
\bibitem[2016b]{RN} Rauw, G., \& Naz\'e, Y.\ 2016b, Advances in Space Research, in press, arXiv1509.06480
\bibitem[2006a]{SGR} Sana, H., Gosset, E., \& Rauw, G.\ 2006a, MNRAS, 371, 67
\bibitem[2006b]{Sana} Sana, H., Rauw, G., Naz\'e, Y., Gosset, E., \& Vreux, J.-M.\ 2006b, MNRAS, 372, 661
\bibitem[2011]{HD93250} Sana, H., Le Bouquin, J.-B., De Becker, M., Berger, J.-P., de Koter, A., \& M\'erand, A.\ 2011, ApJ, 740, L43
\bibitem[2014]{VLTI} Sana, H., Le Bouquin, J.-B., Lacour, S., et al.\ 2014, ApJS, 215, 15
\bibitem[1985]{Shull} Shull, J.M., \& Van Steenberg, M.E.\ 1985, ApJ, 294, 599
\bibitem[2001]{apec} Smith, R.K., \& Brickhouse, N.S.\ 2001, ApJ, 556, L91
\bibitem[1992]{SBP} Stevens, I.R., Blondin, J.M., \& Pollock, A.M.T.\ 1992, ApJ, 386, 265
\bibitem[1994]{SP} Stevens, I.R., \& Pollock, A.M.T.\ 1994, MNRAS, 269, 226
\bibitem[2001]{pn} Str\"uder, L., Briel, U., Dennerl, K., et al.\ 2001, A\&A, 365, L18
\bibitem[2000]{Sung} Sung, H., Chun, M.-Y., \& Bessell, M.S.\ 2000, AJ, 120, 333
\bibitem[2001]{MOS} Turner, M.J.L., Abbey, A., Arnaud, M., et al.\ 2001, A\&A, 365, L27
\bibitem[2008]{Sven} van Loo, S., Blomme, R., Dougherty, S.M., \& Runacres, M.C.\ 2008, A\&A, 483, 585
\bibitem[1992]{Usov} Usov, V.V.\ 1992, ApJ, 389, 635
\bibitem[2001]{Vink} Vink, J.S., de Koter, A., \& Lamers, H.J.G.L.M.\ 2001, A\&A, 369, 574
\bibitem[1998]{Yan} Yan, M., Sadeghpour, H.R., \& Dalgarno, A.\ 1998, ApJ, 496, 1044
\bibitem[1967]{WHS} Wolfe, R.H.Jr., Horak, H.G., \& Storer, N.W.\ 1965, in Modern Astrophysics. A Memorial to Otto Struve, ed.\ M.\ Hack, (New York: Gordon \& Breach), 251
\bibitem[2007]{Zhekov} Zhekov, S.A.\ 2007, MNRAS, 382, 886
\bibitem[2000]{ZS} Zhekov, S.A., \& Skinner, S.L.\ 2000, ApJ, 538, 808
\end{thebibliography}
\end{document}